\documentclass[aos,MSNbibl,seceqn,dvips]{arximspdf}
\usepackage{algorithmic,algorithm}
\usepackage{graphicx}

\doi{10.1214/12-AOS1034} %kopijuoti is PTS
\volume{40}
\issue{4}
\pubyear{2012}
\firstpage{2195}
\lastpage{2238}

\makeatletter
\newcommand{\eqref}[1]{(\ref{#1})}
\newtheorem{theorem}{Theorem}[section]
\newtheorem{lemma}[theorem]{Lemma}
\newproclaim{definition}[theorem]{Definition}
\newcommand{\opnorm}[1]{\|#1\|}
\newcommand{\fronorm}[1]{\|#1\|_{F}}
\newcommand{\onenorm}[1]{\|#1\|_{\ell_1}}
\newcommand{\twonorm}[1]{\|#1\|_{\ell_2}}
\newcommand{\infnorm}[1]{\|#1\|_{\ell_\infty}}
\newcommand{\abs}[1]{|#1|}
\newcommand{\R}{\mathbb{R}}
\newcommand{\sgn}[1]{\operatorname{sgn}(#1)}
\newcommand{\mct}[1]{\bolds{#1}}
\newcommand{\vct}[1]{\mathbf{#1}}
\newcommand{\mtx}[1]{\mathbf{#1}}
\makeatother

\begin{document}
\begin{frontmatter}

\title{A geometric analysis of subspace clustering with~outliers}
\runtitle{Subspace clustering with outliers}

\begin{aug}
\author[A]{\fnms{Mahdi} \snm{Soltanolkotabi}\ead[label=e1]{mahdisol@stanford.edu}\thanksref{t1}}
\and
\author[A]{\fnms{Emmanuel J.} \snm{Cand\'es}\corref{}\ead[label=e2]{candes@stanford.edu}\thanksref{t2}}
\thankstext{t1}{Supported by a Benchmark Stanford Graduate Fellowship.}
\thankstext{t2}{Supported in part by NSF via Grants CCF-0963835,
CNS-0911041 and the 2006 Waterman Award, by AFOSR under Grant
FA9550-09-1-0643, and by ONR under Grant N00014-09-1-0258.}
\runauthor{M. Soltanolkotabi and E. J. Cand\'es}
\affiliation{Stanford University}
\address[A]{Department of Electrical Engineering\\
Stanford University\\
350 Serra Mall\\
Stanford California, 94305\\
USA\\
\printead{e1}\\
\phantom{E-mail:\ }\printead*{e2}} %adresu isvedimo komanda gale!
\end{aug}

% HISTORY:
\received{\smonth{1} \syear{2012}}
\revised{\smonth{7} \syear{2012}}

% ABSTRACT
%
\begin{abstract}
This paper considers the problem of clustering a collection of
unlabeled data points assumed to lie near a union of lower-dimensional
planes. As is common in computer vision or unsupervised
learning applications, we do not know in advance how many subspaces
there are nor do we have any information about their dimensions. We
develop a novel geometric analysis of an algorithm named \textit{sparse
subspace clustering} (SSC) [In \textit{IEEE Conference on Computer Vision and Pattern Recognition, 2009. CVPR 2009}
(2009) 2790--2797. IEEE], which significantly
broadens the range of problems where it is provably effective. For
instance, we show that SSC can recover multiple subspaces, each of
dimension comparable to the ambient dimension. We also prove that
SSC can correctly cluster data points even when the subspaces of
interest intersect. Further, we develop an extension of SSC that
succeeds when the data set is corrupted with possibly overwhelmingly
many outliers. Underlying our analysis are clear geometric insights,
which may bear on other sparse recovery problems. A~numerical study
complements our theoretical analysis and demonstrates the
effectiveness of these methods.
\end{abstract}

% KEYWORDS

\end{frontmatter}

%s1 #&#
\section{Introduction}
\label{secintro}
%s1.1 #&#
\subsection{Motivation}
\label{secmotivation}

One of the most fundamental steps in data analysis and dimensionality
reduction consists of approximating a given data set by a \textit{single}
low-dimensional subspace, which is classically achieved via Principal
Component Analysis (PCA). % More recent results focus on approximating a
% single low dimensional subspace corrupted by sparse noise
% \cite{RPCA}.
%
%f1 #&#
\begin{figure}

\includegraphics{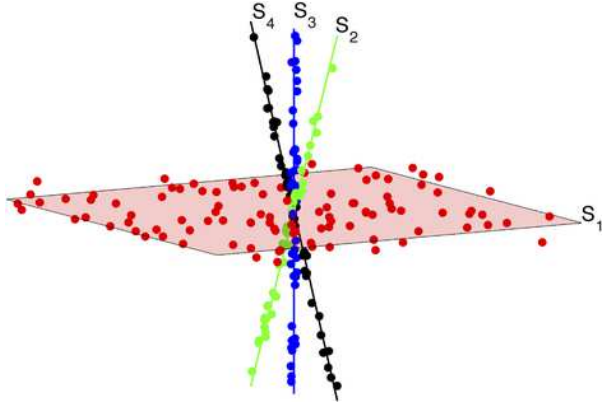}

\caption{Collection of points near a union of multiple subspaces.}\label{figsubspaces}
\end{figure}
In many problems, however, a collection of points may not lie near a
low-dimensional plane but near a union of \textit{multiple} subspaces as
shown in Figure~\ref{figsubspaces}. It is then of interest to find or
fit all these subspaces. Furthermore, because our data points are
unlabeled in the sense that we do not know in advance to which
subspace they belong to, we need to simultaneously cluster these data
into multiple subspaces \textit{and} find a low-dimensional subspace
approximating all the points in a cluster. This problem is known as
\textit{subspace clustering} and has numerous applications; we list just
a few:
\begin{itemize}
%``A gene expression data set from a microarray experiment can
%be represented by a real-valued expression matrix, where the rows form
%the expression pattern of genes, the columns represent the expression
%profiles of samples, and each entry is the measured expression level
%of the corresponding gene and sample.
%Both the gene-based and sample-based clustering approaches search
%exclusive and exhaustive
%partitions of objects that share the same feature space. However,
%current thinking in molecular
%biology holds that only a small subset of genes participate in any
%cellular process of interest and
%that a cellular process takes place only in a subset of the samples.
%This belief calls for the subspace
%clustering to capture clusters formed by a subset of genes across a
%subset of samples. For subspace
%clustering algorithms, genes and samples are treated symmetrically, so
%that either genes or samples
%can be regarded as objects or features. Furthermore, clusters
%generated through such algorithms may
%have different feature spaces."
%
\item\textit{Unsupervised learning}. In unsupervised learning the goal
is to build representations of machine inputs, which can be used for
decision making, predicting future inputs, efficiently communicating
the inputs to another machine and so on.

In some unsupervised learning applications, the standard assumption
is that the data is well approximated by a union of lower-dimensional
manifolds. Furthermore, these manifolds are sometimes
well approximated by subspaces whose dimension is only slightly
higher than that of the manifold under study. Such an example is
handwritten digits. When looking at handwritten characters for
recognition, the human eye is able to allow for simple
transformations such as rotations, small scalings, location shifts
and character thickness. Therefore, any reasonable model should be
insensitive to such changes as well. Simard et al. \cite{Simard93}
characterize this invariance with a $7$-dimensional manifold; that
is, different transformations of a single digit are well
approximated by a $7$-dimensional manifold. As illustrated by Hastie
et al. \cite{Hastie97}, these $7$-dimensional manifolds are in turn
well approximated by $12$-dimensional subspaces. Thus, in certain
cases, unsupervised learning can be formulated as a subspace
clustering problem.

%Another learning problem where subspace clustering is relevant is the
%problem of Multi-label Classification. \MS{I have to discuss this with
%you before including it}

\item\textit{Computer vision}. There has been an explosion of visual
data in the past few years. Cameras are now everywhere: street
corners, traffic lights, airports and so on. Furthermore, millions
of videos and images are uploaded monthly on the web. This visual
data deluge has motivated the development of low-dimensional
representations based on appearance, geometry and dynamics of a
scene. In many such applications, the low-dimensional
representations are characterized by multiple low-dimensional
subspaces. One such example is motion segmentation
\cite{Vidal08}. Here, we have a video sequence which consists of
multiple moving objects, and the goal is to segment the trajectories
of the objects. Each trajectory approximately lies in a
low-dimensional subspace. To understand scene dynamics, one needs to
cluster the trajectories of points on moving objects based on the
subspaces (objects) they belong to, hence the need for subspace
clustering.

Other applications of subspace clustering in computer vision include
image segmentation \cite{Yang08}, face clustering \cite{Ho03}, image
representation and compression~\cite{Hong06}, and systems theory~\cite{Vidal03}.
Over the years, various methods for subspace
clustering have been proposed by researchers working in this
area. For a comprehensive review and comparison of these algorithms,
we refer the reader to the tutorial \cite{Vidal11} and references
therein
\cite{Boult91,Costeira98,Gear98,Vidal05,Bradley00,Tseng00,Agarwal04,Lu06,Zhang09,Tipping99,Sugaya04,Ma07,Rao08,Yang06,Yan06,Zhang10,Goh07,Elhamifar09,Elhamifar10,Liu10,Chen09}.
\item\textit{Disease detection}. In order to detect a class of diseases
of a specific kind (e.g., metabolic), doctors screen specific factors
(e.g., metabolites). For this purpose, various tests (e.g., blood
tests) are performed on the newborns and the level of those
factors are measured. One can further construct a newborn-factor
level matrix, where each row contains the factor levels of a
different newborn. That is to say, each newborn is associated with a
vector containing the values of the factors. Doctors wish to
cluster groups of newborns based on the disease they suffer
from. Usually, each disease causes a correlation between a specific
set of factors. Such an assumption implies that points corresponding
to newborns suffering from a given disease lie on a lower-dimensional
subspace \cite{Peterkriegel08}. Therefore, the
clustering of newborns based on their specific disease together with
the identification of the relevant factors associated with each
disease can be modeled as a subspace clustering problem.
\end{itemize}

PCA is perhaps the single most important tool for dimensionality
reduction. However, in many problems, the data set under study is not
well approximated by a linear subspace of lower dimension. Instead,
as we hope we have made clear, the data often lie near a union of
low-dimensional subspaces, reflecting the multiple categories or
classes a set of observations may belong to. Given its relevance in
data analysis, we find it surprising that subspace clustering has been
well studied in the computer science literature but has comparably
received little attention from the statistical community. This paper
begins with a very recent approach to subspace clustering and
proposes a framework in which one can develop some useful statistical
theory. As we shall see, insights from sparse regression analysis in
high dimensions---a subject that has been well developed in the
statistics literature in recent years---inform the subspace clustering
problem.

% In this paper
% we focus on a recently developed algorithm (SSC) and its theoretical
% analysis; highlighting statistical aspects of the problem.

%s1.2 #&#
\subsection{Problem formulation}
\label{secModel}

In this paper we assume we are given data points that are distributed
on a union of unknown linear subspaces $S_1 \cup S_2 \cup\cdots
\cup S_L$; that is, there are $L$ subspaces of $\mathbb{R}^n$ of
unknown dimensions $d_1,d_2,\ldots, d_L$. More precisely, we have a
point set
$\mathcal{X}\subset\mathbb{R}^n$ consisting of $N$ points in
$\mathbb{R}^n$, which may be partitioned as
%
%e1.1 #&#
\begin{equation}
\label{equnion} \mathcal{X} = \mathcal{X}_0 \cup
\mathcal{X}_1 \cup\cdots\cup \mathcal{X}_L
\end{equation}
for each $\ell\ge1$, $\mathcal{X}_\ell$ is a collection of $N_\ell$
unit-normed vectors chosen from
$S_\ell$. The careful reader
will notice that we have an extra subset $\mathcal{X}_0$ in
\eqref{equnion} accounting for possible outliers. Unless specified
otherwise, we assume that this special subset consists of $N_0$ points
chosen independently and uniformly at random on the unit sphere. The
task is now simply stated. Without any prior knowledge about the
number of subspaces, their orientation or their dimension,
\begin{longlist}[(1)]
\item[(1)] identify all the outliers, and
\item[(2)] segment or assign each data point to a cluster as to recover
all the hidden subspaces.
\end{longlist}

It is worth emphasizing that our model assumes normalized data vectors;
this is not a restrictive assumption since one can always normalize
inputs before applying any subspace clustering algorithm. Although we
consider linear subspaces, one can extend the methods of this paper to
affine subspace clustering which will be explained in Section \ref
{subsecmethods}.%In any
%event, our results extend to the nonunit norm case with very simple
%modifications.

%In this paper we focus on noiseless data. For subspace recovery in the
%presence of noise we refer the reader to our future work
We now turn to methods for achieving these goals. Our focus is on
noiseless data and we
leave noisy subspace clustering to future work.
%refer the reader to \cite{Elhamifar11} for work
%concerning subspace recovery from noisy samples.
%s1.3 #&#
\subsection{Methods and contributions}

To introduce our methods, we first consider the case in which there
are no outliers before treating the more general case. From now on, it
will be convenient to arrange the observed data points as columns of a
matrix $\mtx{X} = [\vct{x}_1, \ldots, \vct{x}_N] \in
\mathbb{R}^{n\times N}$, where $N = N_0 + N_1 + \cdots+ N_L$ is the
total number of points.

%s1.3.1 #&#
\subsubsection{Methods}
\label{subsecmethods}
Subspace clustering has received quite a bit of attention in recent
years and, in particular, Elhamifar and Vidal
introduced a clever algorithm based on insights from the \textit{compressive sensing}
literature. The key idea of the Sparse Subspace
Clustering (SSC) algorithm \cite{Elhamifar09} is to find the \textit{sparsest}
expansion of each column $\vct{x}_i$ of $\mtx{X}$ as a
linear combination of all the other columns. This makes a lot of sense
because under some generic conditions,
% ($\hat{\mathbf{X}}_i=[\mathbf{x}_1,\ldots,\mathbf{x}_{i-1},
% \ldots,\mathbf{x}_{N}]$).
one expects that the \textit{sparsest} representation of $\vct{x}_i$\vadjust{\goodbreak}
would only select vectors from the subspace in which~$\vct{x}_i$
happens to lie in. This motivates Elhamifar and Vidal to consider the
sequence of optimization problems
%
%e1.2 #&#
\begin{equation}
\label{eqel2} \min_{\vct{z} \in\R^N} \|\vct{z}\|_{\ell_1} \qquad\mbox{subject
to } \mtx{X} \vct{z} = \vct{x}_i \mbox{ and } z_i=0.
\end{equation}
The hope is that whenever $z_j \neq0$, $\vct{x}_i$ and $\vct{x}_j$
belong to the same subspace. This property is captured by the
definition below.
%
%de1.1 #&#
\begin{definition}[($\ell_1$ subspace detection property)] The subspaces
$\{S_\ell\}_{\ell=1}^{L}$ and points $\mtx{X}$ obey the $\ell_1$
subspace detection property if and only if it holds that for all~$i$,
the optimal solution to \eqref{eqel2} has nonzero entries only
when the corresponding columns of $\mtx{X}$ are in the same
subspace as $\vct{x}_i$.
\end{definition}

\newcommand{\sdp}{subspace detection property}

%It is known that if the subspaces are independent, then the \sdp holds
%bound concerning the minimum principal angle --- $\min_{k \neq\ell}
%purpose of this paper is to extend this line of work and show that the
%by using a statistical framework.
In certain cases the \sdp\ may not hold, that is, the support of the
optimal solution to \eqref{eqel2} may include points from other
subspaces. However, it might still be possible to detect and construct
reliable clusters. A strategy is to arrange the optimal solutions to
\eqref{eqel2} as columns of a matrix $\mtx{Z}\in\mathbb{R}^{N\times
N}$, build an affinity graph $G$ with $N$ vertices and weights
$w_{ij}=\abs{Z_{ij}}+\abs{Z_{ji}}$, construct the normalized Laplacian
of $G$, and use a gap in the distribution of eigenvalues of this
matrix to estimate the number of subspaces. Using the estimated number
of subspaces, spectral clustering techniques (e.g., \cite{Shi00,Ng02})
can be applied to the affinity graph to cluster the data points. The
main steps of this procedure are summarized in Algorithm
\ref{algSSC}. This algorithm clusters linear subspaces but can also
cluster affine subspaces by adding the constraint
$\mtx{Z}^T\vct{1}=\vct{1}$ to \eqref{eqel2}.

\begin{algorithm}
\caption{Sparse subspace clustering (SSC)}
\begin{algorithmic}
\REQUIRE{A data set $\mathcal{X}$ arranged as columns of
$\mtx{X}\in\mathbb{R}^{n\times N}$.} % from a union of $L$ subspaces
% $\{S_\ell\}_{\ell=1}^L$.}
\STATE1. Solve (the optimization variable is the $N \times N$
matrix $\mtx{Z}$)
\begin{eqnarray*}
&&\mbox{minimize}\qquad \|\mtx{Z}\|_{\ell_1}
\\
&&\mbox{subject to} \qquad \mtx{X}\mtx{Z} = \mtx{X},
\\
&&\hphantom{\mbox{subject to} \qquad} \operatorname{diag}(\mtx{Z}) = \vct{0}.
\end{eqnarray*}
\STATE2. Form the affinity graph $G$ with nodes representing the $N$
data points and edge weights given by
$\mtx{W}=\abs{\mtx{Z}}+\abs{\mtx{Z}}^T$. \STATE3. Sort the
eigenvalues $\sigma_1 \ge\sigma_2 \ge\cdots\ge\sigma_N$ of the
normalized Laplacian of $G$ in descending order, and set
\[
\hat{L}=N-\mathop{\arg\max}_{i=1,\ldots,N-1} (\sigma_i-
\sigma_{i+1}).
\]
\STATE4. Apply a spectral clustering
technique to the affinity graph using $\hat{L}$ as the estimated
number of clusters. \ENSURE{Partition
$\mathcal{X}_1,\ldots,\mathcal{X}_{\hat{L}}$.}
\end{algorithmic}
\label{algSSC}
\end{algorithm}
%
%s1.3.2 #&#
\subsubsection{Our contributions}
In Section~\ref{comparison} we will review existing conditions
involving a restriction on the minimum angle between subspaces under
which Algorithm~\ref{algSSC} is expected to work. The main purpose of
this paper is to show that Algorithm~\ref{algSSC} works in much
broader situations.
\begin{itemize}
\item\textit{Subspaces with nontrivial intersections.} Perhaps
unexpectedly, we shall see that our results assert that SSC can
correctly cluster data points even when our subspaces intersect so
that the minimum principal angle vanishes. This is a phenomenon
which is far from being explained by current theory.

\item\textit{Subspaces of nearly linear dimension.} We prove that
in generic settings, SSC can effectively cluster the data even when
the dimensions of the subspaces grow almost linearly with the ambient
dimension. We are not aware of other literature explaining why this
should be so. To be sure, in most favorable cases, earlier results
only seem to allow the dimensions of the subspaces to grow at most
like the square root of the ambient dimension.

\item\textit{Outlier detection.} We present modifications to SSC that
succeed when the data set is corrupted with many outliers---even
when their number far exceeds the total number of clean
observations. To the best\vadjust{\goodbreak} of our knowledge, this is the first
algorithm provably capable of handling these many corruptions.

\item\textit{Geometric insights.} Such improvements are possible
because of a novel approach to analyzing the sparse subspace
clustering problem. This analysis combines tools from convex
optimization, probability theory and geometric functional
analysis. Underlying our methods are clear geometric insights
explaining quite precisely when SSC is successful and when it is
not. This viewpoint might prove fruitful to address other sparse
recovery problems.
\end{itemize}
Section~\ref{comparison} proposes a careful comparison with the
existing literature. Before doing so, we first need to introduce our
results, which is the object of Sections~\ref{typical} and
\ref{MainResults}.

%s1.4 #&#
\subsection{Models and typical results}
\label{typical}

%s1.4.1 #&#
\subsubsection{Models}
\label{subsecformodel}

In order to better understand the regime in which SSC succeeds as well
as its limitations, we will consider three different models. Our aim
is to give informative bounds for these models highlighting the
dependence upon key parameters of the problem such as (1) the number of
subspaces, (2) the dimensions of these subspaces, (3) the relative
orientations of these subspaces, (4) the number of data points per
subspace and so on.\looseness=1
\begin{itemize}
\item\textit{Deterministic model.} In this model the orientation of
the subspaces as well as the distribution of the points on each
subspace are nonrandom. This is the setting considered by Elhamifar
et al. and is the subject of Theorem~\ref{th1}, which guarantees
that the subspace detection property holds as long as for any two
subspaces, pairs of (primal and dual) directions taken on each
subspace have a sufficiently small inner product.
% between precise directions on one
% subspace and certain directions on the other subspace are small
% enough.}
%
\item\textit{Semi-random model.} Here, the subspaces are fixed but
the points are distributed at random on each of the subspaces. This
is the subject of Theorem~\ref{th2}, which uses a notion of affinity
to measure closeness between any two subspaces. This affinity is
maximal and equal to the square root of the dimension of the
subspaces when they overlap perfectly. Here, our results state that
if the affinity is smaller, by a logarithmic factor, than its
maximum possible value, then SSC recovers the subspaces
exactly.
% affinity
% between subspaces. The affinity measures the closeness between two
% subspaces and is maximal when they
% is smaller than a constant (up to log factors) times the
% highest possible affinity(square root of the dimension of the
% subspaces), SSC will recover the subspaces exactly.

\item\textit{Fully random model.} Here, both the orientation of the
subspaces and the distribution of the points are random. This is the
subject of Theorem~\ref{th0}; in a nutshell, SSC succeds as long as
the dimensions of the subspaces are within at most a logarithmic
factor from the ambient dimension.
\end{itemize}

%s1.4.2 #&#
\subsubsection{Segmentation without outliers}

Consider the fully random model first. We establish that the
\sdp\ holds as long as the dimensions of the subspaces are
roughly linear in the ambient dimension. Put differently, SSC can
provably achieve perfect subspace recovery in settings not previously
understood.

Our results make use of a constant $c(\rho)$ only depending upon the
density of inliers (the number of points on each subspace is $\rho
d+1$) and which obeys the following two properties:
\begin{longlist}[(ii)]
\item[(i)] For all $\rho> 1$, $c(\rho)>0$.
\item[(ii)] There is a numerical value $\rho_0$, such that for all
$\rho\ge\rho_0$, one can take $c(\rho)=\frac{1}{\sqrt{8}}$.
\end{longlist}
%
%th1.2 #&#
\begin{theorem}
\label{th0}
Assume there are $L$ subspaces, each of dimension $d$, chosen
independently and uniformly at random. Furthermore, suppose there are
$\rho d+1$ points chosen independently and uniformly at random on each
subspace.\setcounter{footnote}{2}\footnote{From here on, when we say that points are chosen
from a subspace, we implicitly assume they are unit normed. For ease
of presentation we state our results for $1<\rho\le
e^{{d}/{2}}$, that is, the number of points on each subspace is not
exponentially large in terms of the dimension of that subspace. The
results hold for all $\rho>1$ by replacing $\rho$ with
$\min\{\rho,e^{{d}/{2}}\}$.} Then the subspace detection
property holds with large probability as long as
%
%e1.3 #&#
\begin{equation}
\label{lincond} d< \frac{c^2(\rho) \log\rho}{12\log N} n
\end{equation}
[$N=L(\rho d+1)$ is the total number of data points]. The probability
is at least $1-\frac{2}{N}-Ne^{-\sqrt{\rho}d}$, which is calculated
for values of $d$ close to the upper bound. For lower values of $d$,
the probability of success is of course much higher, as explained
below.
\end{theorem}
Theorem~\ref{th0} is in fact a special instance of a more general
theorem that we shall discuss later and which holds under less
restrictive assumptions on the orientations of the subspaces as well
as the number and positions of the data points on each subspace. This
theorem conforms to our intuition since clustering becomes more
difficult as the dimensions of the subspaces increase. Intuitively,
another difficult regime concerns a situation in which we have very
many subspaces of small dimensions. This difficulty is reflected in
the dependence of the denominator in \eqref{lincond} on $L$, the
number of subspaces (through $N$). A more comprehensive explanation of
this effect is provided in Section~\ref{semirandomsec}.

As it becomes clear in the proof (see Section~\ref{proofs}), a
slightly more general version of Theorem~\ref{th0} holds, namely, with
$0<\beta\le1$, the subspace detection property holds as long as
%
%e1.4 #&#
\begin{equation}
\label{fullrandwithbeta} d<2\beta \biggl[\frac{c^2(\rho) \log\rho}{12\log N} \biggr] n
\end{equation}
with probability at least
$1-\frac{2}{N}-Ne^{-\rho^{(1-\beta)}d}$. Therefore, if $d$ is a
small fraction of the right-hand side in \eqref{lincond}, the
\sdp\ holds with much higher probability, as expected.

%An interesting regime is when the number of subspaces $L$ is fixed and
%$N=n^\gamma$, with $\gamma>1$. Then as $n\rightarrow\infty$, it
%follows from $N \asymp L \rho d$ that our theorem implies that the
%d<c   n
%for some numerical constant $c$. This justifies our earlier claims
%since we can have subspace dimensions growing linearly in the ambient
%dimension.

% \EJC{The suggestion is better so we should go for it. I would change
% it to say that the number of points is polynomial in $d$ and that
% the total number of points is polynomial in the ambient
% dimension. You also need to be careful because you overload the
% symbol $\gamma$ again. If we have a result with 48, then why not
% state it directly? We need to keep the sentence about the close gap
% between theory and practice. }
% \MS{from here to the beginning of the next paragraph is the new
%version. Also please check that you agree with the corresponding
%footnote. In this format there is not a conflict of notation as in the
%previous case.}

An interesting regime is when the number of subspaces $L$ is fixed and
the density of points per subspace is $\rho=d^\eta$, for a small
$\eta>0$. Then as $n\rightarrow\infty$ with the ratio $d/n$
fixed, it follows from $N \asymp L \rho d$ and
(\ref{fullrandwithbeta}) using $\beta=1$ that the \sdp\ holds as long
as
\[
d<\frac{\eta}{48(1+\eta)} n.
\]
This justifies our earlier claims
since we can have subspace dimensions growing linearly in the ambient
dimension. It should be noted that this asymptotic statement is only a
factor $8-10$ away from what is observed in simulations, which
demonstrates a relatively small gap between our theoretical
predictions and simulations.\footnote{To be concrete, when the ambient
dimension is $n=50$ and the number of subspaces is $L=10$, the
subspace detection property holds for $d$ in the range from $7$ to
$10$.} % \EJC{This is a placeholder to remind ourselves that the
% dependence on $\eta$ above is funny.}\MS{I agree that this does not
% look good, in order to avoid confusion for the reader (so that they
% do not actually try to plug in a number for $\eta$) I suggest adding
% the adjective small before $\eta$}
% \EJC{If we plug in $\beta= 1$ in the prob. after (1.4), we have a
% problem. I am also not sure I understand why having more points per
% subspace makes the problem harder.}
% \MS{This is for when the ratio is fixed, so as $n$ goes to infinity
%$d$ also goes to infinity. You are correct about the dependence on the
%number of points. The problem is that the denominator, I think should
%really be $\log\frac{N}{n}$. However, our analysis is not capable of
%getting this. There was a $\log L$ factor missing which I've included.}

%s1.4.3 #&#
\subsubsection{Segmentation with outliers}

We now turn our attention to the case where there our extraneous points
in the data
in the sense that there are $N_0$ outliers assumed to be distributed
uniformly at random on the unit sphere. Here, we wish to correctly
identify the outlier points and apply any of the subspace clustering
algorithms to the remaining samples.\vadjust{\goodbreak} We propose a very simple
detection procedure for this task. As in SSC, decompose each
$\vct{x}_i$ as a linear combination of all the other points by solving
an $\ell_1$-minimization problem. Then one expects the expansion of an
outlier to be less sparse. This suggests the following detection rule:
declare $\vct{x}_i$ to be an outlier if and only if the optimal value
of \eqref{eqel2} is above a fixed threshold. This makes sense because
if $\vct{x}_i$ is an outlier, one expects the optimal value to be on
the order of $\sqrt{n}$ (provided $N$ is at most polynomial in $n$),
whereas this value will be at most on the order of $\sqrt{d}$ if
$\vct{x}_i$ belongs to a subspace of dimension $d$. In short, we
expect a gap---a fact we will make rigorous in the next section.
% As long as the number of points are not exponential in terms of the
% ambient dimension ($n$), when $\mathbf{x}_i$ is an outlier point
% chosen uniformly at random from the unit sphere, one expects the
% optimal value of~\ref{eqel3} to be around $\sqrt{n}$. Intuitively,
% the reason is that to be able to approximate the unit sphere well
% you need exponentially many points. Similarly, when $\mathbf{x}_i$
% is a data point residing in a $d$ dimensional subspace, one expects
% that the optimal value of~\ref{eqel3} to be around $\sqrt{d}$. This
% suggests that one could use the optimal value of~\ref{eqel3} as a
% detection criterion. In other words as long as
% $\max_\ell\sqrt{d_\ell}<\sqrt{n}$ one expects a gap to appear in
% optimal values of~\ref{eqel3} i.e. the optimal value is small when
% $\mathbf{x}_i$ resides on a subspace and large when $\mathbf{x}_i$
% is an outlier point. We will formalize this intuition in the next
% section.
The main steps of the procedure are shown in Algorithm
\ref{algoutlier}.
\begin{algorithm}[t]
\caption{Subspace clustering in the presence of outliers}
\begin{algorithmic}
\REQUIRE{A data set $\mathcal{X}$ arranged as columns of
$\mtx{X}\in\mathbb{R}^{n\times N}$. % from union of $L$ subspaces
% $\{S_\ell\}_{\ell=1}^L$ and $N_0$ outliers chosen uniformly at
% random from the unit sphere.
}
\STATE1. Solve
\begin{eqnarray*}
&&\mbox{minimize}\qquad  \|\mtx{Z}\|_{\ell_1}
\\
&&\mbox{subject to}\qquad  \mtx{X}\mtx{Z} = \mtx{X},
\\
&&\phantom{\mbox{subject to}\qquad} \operatorname{diag}(\mtx{Z}) = \vct{0}.
\end{eqnarray*}
\STATE2. For each $i \in\{1, \ldots, N\}$, declare $i$ to be an
outlier iff $\onenorm{\vct{z}_i}>
\lambda(\gamma)\sqrt{n}$.\footnotemark[5]
\STATE3. Apply a subspace clustering to the remaining points.
\ENSURE{Partition $\mathcal{X}_0, \mathcal{X}_1, \ldots,
\mathcal{X}_L$.}
\end{algorithmic}
\label{algoutlier}
\end{algorithm}

Our second result asserts that as long as the number of outliers is
not overwhelming, Algorithm~\ref{algoutlier} detects all of them.\footnotetext[5]{Here, $\gamma=\frac{N-1}{n}$ is the
\textit{total point density} and $\lambda$ is a threshold ratio function
whose value shall be discussed later.}\vspace*{-3pt}
%
%th1.3 #&#
\begin{theorem}
\label{th0outlier}
Assume there are $N_{d}$ points to be clustered together with $N_0$
outliers sampled uniformly at random on the $n-1$-dimensional unit
sphere ($N = N_0 + N_d$). Algorithm~\ref{algoutlier} detects all of
the outliers with high probability\setcounter{footnote}{5}\footnote{With probability at least
$1-N_0e^{-Cn/\log(N_0+N_d)}$. If
$N_0<\frac{1}{n}e^{c\sqrt{n}}-N_d$,
this is at least $1-\frac{1}{n}$.} as long as%
%
%e1.5 #&#
\[
\label{ouroutliers1} N_0<\frac{1}{n}e^{c\sqrt{n}}-N_d,
\]
where $c$ is a numerical constant. Furthermore, suppose the subspaces
are $d$-dimensional and of arbitrary orientation, and that each
contains $\rho d+1$ points sampled independently and uniformly at
random. Then with high probability,\footnote{With probability at least
$1-N_0e^{-Cn/\log(N_0+N_d)}-N_de^{-\sqrt{\rho}d}$. If
$N_0<\min\{ne^{c_2{n}/{d}},\break\frac{1}{n}e^{c\sqrt{n}}\}-N_d$, this is
at least $1-\frac{1}{n}-N_de^{-\sqrt{\rho}d}$.} Algorithm\vadjust{\goodbreak}
\ref{algoutlier} does not detect any subspace point as outlier
provided that
%
%e1.6 #&#
\[
\label{ouroutliers2} N_0<n\rho^{c_2{n}/{d}}-N_d ,
\]
in which $c_2={c^2(\rho)}/{(2e^2\pi)}$.
\end{theorem}

This result shows that our outlier detection scheme can reliably
detect all outliers even when their number grows exponentially in the
root of the ambient dimension. We emphasize that this holds without
making any assumption whatsoever about the orientation of the
subspaces or the distribution of the points on each
subspace. Furthermore, if the points on each subspace are uniformly
distributed, our scheme will not wrongfully detect a subspace point as
an outlier. In the next section we show that similar results hold
under less restrictive assumptions.

%s2 #&#
\section{Main results}
\label{MainResults}

%s2.1 #&#
\subsection{Segmentation without outliers}\label{WOth}

In this section we shall give sufficient conditions in the fully
deterministic and semi-random model under which the SSC algorithm
succeeds (we studied the fully random model in Theorem~\ref{th0}).

Before we explain our results, we introduce some basic notation. We
will arrange the $N_\ell$ points on subspace
$S_\ell$ as columns of a matrix $\mtx{X}^{(\ell)}$. For
$\ell=1,\ldots,L$, $i=1,\ldots,N_\ell$, we use $\mtx{X}^{(\ell)}_{-i}$
to denote all points on subspace $S_\ell$ excluding the $i$th point,
$\mtx{X}^{(\ell)}_{-i}=[\vct{x}^{(\ell)}_1,\ldots,\vct{x}^{(\ell
)}_{i-1},\vct{x}^{(\ell)}_{i+1},\ldots,\vct{x}^{(\ell)}_{N_\ell}]$. We
use $\mtx{U}^{(\ell)}\in\R^{n\times d_\ell}$ to denote an arbitrary
orthonormal basis for $S_\ell$. This induces a factorization
$\mtx{X}^{(\ell)}=\mtx{U}^{(\ell)}\mtx{A}^{(\ell)}$, where
$\mtx{A}^{(\ell)}=
[\matrix{\vct{a}_1^{(\ell)},\ldots,\vct{a}_{N_\ell}^{(\ell)}
}]
\in\R^{d_\ell\times
N_\ell}$ is a matrix of coordinates with unit-norm columns. For any
matrix $\mtx{X}\in\mathbb{R}^{n \times N}$, the shorthand notation
$\mathcal{P}(\mtx{X})$ denotes the symmetrized convex hull of its
columns,
$\mathcal{P}(\mtx{X})=\operatorname{conv}(\pm\vct{x}_1,\pm\vct
{x}_2,\ldots,\pm\vct{x}_N)$. Also
$\mathcal{P}^\ell_{-i}$ stands for
$\mathcal{P}(\mtx{X}^{(\ell)}_{-i})$. Finally, $\opnorm{\mtx{X}}$ is
the operator norm of $\mtx{X}$ and $\infnorm{\mtx{X}}$ the maximum
absolute value of its entries.

%s2.1.1 #&#
\subsubsection{Deterministic model}
% the two parts of This is the model where everything is
% deterministic; that is both the orientation of the subspaces as well
% as the distribution of the points on the subspaces are
% nonrandom. In this setting Theorem~\ref{th1} guarantees that the
% $\ell_1$ Subspace Detection property holds; i.e. the matrix
% $\mtx{Z}$ is a block diagonal matrix.\footnote{From here on when we
% say block diagonal we implicitly mean a permutation of the block
% diagonal matrix} Notice that the goal is to give a sufficient
% condition that one is able to compute in at least a probabilistic
% setting, in fact in the process of the proof of this theorem we give
% an ``almost" necessary and sufficient condition is
% provided.%The first part guarantees that the $\ell_1$ Subspace
%Detection property holds; i.e. the matrix $\mathbf{Z}$ is a block
%diagonal matrix\footnote{From here on when we say block diagonal we
%implicitly mean a permutation of the block diagonal matrix} . The
%second part of the theorem guarantees that the affinity graph of each
%block is connected, therefore the spectral clustering methods will
%succeed.
We first introduce some basic concepts needed to state our
deterministic result.

\begin{figure}

\includegraphics{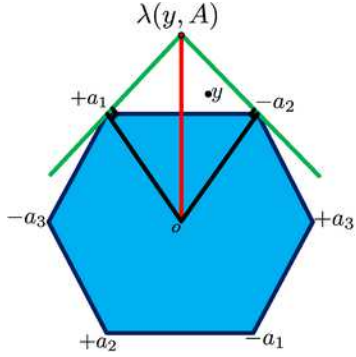}

\caption{Geometric representation of a dual point; see Definition
\protect\ref{dualpointdef}.}\label{figdualpoint}%\vspace*{-3pt}
\end{figure}

%de2.1 #&#
\begin{definition}[(Dual point)]\label{dualpointdef} Consider a vector
$\vct{y}\in\R^d$ and a matrix $\mtx{A}\in\R^{d\times N}$, and let
$\mathcal{C}^*$ be the set of optimal solutions to
%
%e2.1 #&#
\[
\max_{\mct{\lambda} \in\R^d} \langle\vct{y},\mct{\lambda}\rangle \qquad\mbox{subject to
} \bigl\|\mtx{A}^T\mct{\lambda}\bigr\|_{\ell_\infty}\le1.
\]
The dual point $\mct{\lambda}(\vct{y},\mtx{A})\in\R^d$ is defined as a
point in $\mathcal{C}^*$ with minimum Euclidean norm.\footnote{If this
point is not unique, take $\mct{\lambda}(\vct{y},\mtx{A})$ to be any
optimal point with minimum Euclidean norm.} A geometric representation
is shown in Figure
\ref{figdualpoint}.\vadjust{\eject}
% This point can be interpreted as the minimum Euclidean norm point on
% the face of the dual polytope of $\mathcal{P}(\mtx{A})$ which is
% dual to the ``closest face" of $\mathcal{P}(\mtx{A})$ to $\vct{y}$.
\end{definition}
%
%f2 #&#
%
%We will clarify the geometric interpretation of the dual point in
%Section~\ref{Geompers}.
%
%de2.2 #&#
\begin{definition}[(Dual directions)] \label{dualdirectiondef} Define
the dual directions $\vct{v}_i^{(\ell)}\in\R^n$ [arranged as columns
of a matrix $\mtx{V}^{(\ell)}$]
%
%f3 #&#
\begin{figure}[b]

\includegraphics{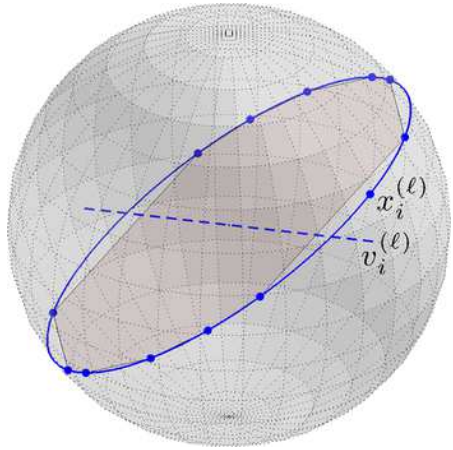}

\caption{Geometric representation of a dual direction. The dual
direction is the dual point embedded in the ambient $n$-dimensional
space.}\label{figdualdirection}%\vspace*{-3pt}
\end{figure}
corresponding to the dual points
$\mct{\lambda}_i^{(\ell)}=\mct{\lambda}(\vct{a}_i^{(\ell)},\mtx
{A}_{-i}^{(\ell)})$
as
\[
\vct{v}^{(\ell)}_i=\mtx{U}^{(\ell)}\frac{\mct{\lambda}_i^{(\ell
)}}{\twonorm{\mct{\lambda}_i^{(\ell)}}}.
\]
The dual direction $ \vct{v}^{(\ell)}_i$, corresponding to the point
$\vct{x}^{(\ell)}_i$, from subspace $S_\ell$ is shown in Figure
\ref{figdualdirection}.
\end{definition}

%
%de2.3 #&#
\begin{definition}[(Inradius)] The inradius of a convex body $\mathcal
{P}$, denoted by $r(\mathcal{P})$, is defined as the radius of the
largest Euclidean ball inscribed in~$\mathcal{P}$.\vadjust{\goodbreak}
\end{definition}

%de2.4 #&#
\begin{definition}[(Subspace incoherence)] The subspace incoherence of a
point set $\mathcal{X}_\ell$ vis a vis the other points is defined
by
\[
\mu(\mathcal{X}_\ell)=\mathop{\max}_{\vct{x}\in\mathcal{X}\setminus\mathcal
{X}_\ell}\bigl \|{
\mtx{V}^{(\ell)}}^T\vct{x}\bigr\|_{\ell_\infty},
\]
where $\mtx{V}^{(\ell)}$ is as in Definition~\ref{dualdirectiondef}.
\end{definition}
%
%Given two subspaces $S_k$ and $S_\ell$ in $\mathbb{R}^n$, with
%corresponding orthonormal basis $\mathbf{V}_k$ and $\mathbf{V}_\ell$,
%we define the canonical distance between the two subspaces by:
%dist(S_k,S_\ell)= \fronorm{\mathbf{V}_k^T\mathbf{V}_\ell}
%notice that this notion of distance is well defined, i.e. the choice
%of the orthonormal basis does not matter because the frobenius norm is
%invariant under orthogonal transforms.
%A set of point in $\mathbb{R}^d$ are said to be in general position if
%, if no $k\le d+1$ points lie in a common $(k-2)$-flat.
%
%th2.5 #&#
\begin{theorem}
\label{th1}
If
%
%e2.2 #&#
\begin{equation}
\label{GeomCond} \mu(\mathcal{X}_\ell)<\mathop{ \min}_{i: \vct{x}_i\in\mathcal{X}_\ell
}
r \bigl(\mathcal{P}^\ell_{-i} \bigr)
\end{equation}
for each $\ell=1,\ldots,L$, then the subspace detection property
holds. If \eqref{GeomCond} holds for a given $\ell$, then a local
subspace detection property holds in the sense that for all~$\vct
{x}_i$, the solution to \eqref{eqel2} has nonzero
entries only when the corresponding columns of $\mtx{X}$ are in the
same subspace as $\vct{x}_i$.
\end{theorem}

%
%f4 #&#
\begin{figure}[b]

\includegraphics{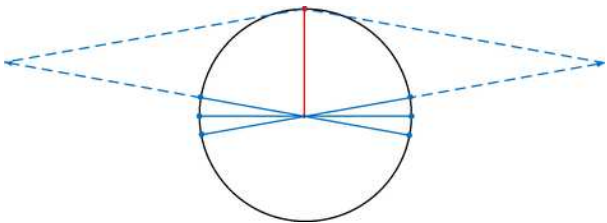}

\caption{Skewed distribution of points on a single subspace and
$\ell_1$ synthesis.}\label{EJCintuition}
\end{figure}

%
%The condition presented above is a sufficient deterministic condition,
%but is not necessary. The proof of this theorem is based on what we
%have called the``modified null space property", which as will be shown
%in the proof is almost necessary and sufficient. However, it is not
%easy to check the condition even in a probabilistic setting.
%This result essentially states that the subspace detection property
%holds as long as, for $\ell=1,\ldots,L$ the points on all subspaces
%except the one on subspace $S_\ell$, have low coherence with the dual
%direction of the points on subspace $S_\ell$.

The incoherence parameter of a set of points on one subspace with
respect to other points is a measure of affinity between subspaces. To
see why, notice that if the incoherence is high, it implies that there
is a point on one subspace and a direction on another (a dual
direction) such that the angle between them is small. That is, there
are two ``close'' subspaces, hence, clustering becomes hard. The
inradius measures the spread of points. A very small minimum inradius
implies that the distribution of points is skewed toward certain
directions, thus, \textit{subspace} clustering using an $\ell_1$ penalty
is difficult. To see why this is so, assume the subspace is of
dimension $2$ and all of the points on the subspace are skewed toward
one line, except for one special point which is in the direction
orthogonal to that line. This is shown in Figure~\ref{EJCintuition}
with the special point in red and the others in blue. To
synthesize this special point as a linear combination of the other
points from its subspace, we would need huge coefficient values and
this is why it may very well be more economical---in an $\ell_1$
sense---to select points from other subspaces. This is a situation
where $\ell_0$ minimization would still be successful but its convex
surrogate is not (researchers familiar with sparse regression would
recognize a setting in which variables are correlated and which is
challenging for the LASSO).
% the $\ell_1$ ``special" point the $\ell_1$ penalty of SSC will not
% pick points from its own subspace. The reason is that once the
% ``special" point is excluded the inradius is very small.
%Intuitively, any subspace clustering strategy will have
%difficulty recognizing whether the points correspond to a subspace of
%dimension $1$ or $2$.
Theorem~\ref{th1} essentially states that as long as
different subspaces are not similarly oriented and the points on a
single subspace are well spread, SSC can cluster the data correctly. A~geometric perspective of \eqref{GeomCond} is provided in Section
\ref{Geompers}.

To get concrete results, one needs to estimate both the incoherence
and inradius in terms of the parameters of interest, which include the
number of subspaces, the dimensions of the subspaces, the number of
points on each subspace and so on. To do this, we use the
probabilistic models we introduced earlier. This is our next
topic.

%s2.1.2 #&#
\subsubsection{Semi-random model}
\label{semirandomsec}
The following definitions capture notions of similarity/affinity
between two subspaces.

%de2.6 #&#
\begin{definition}
The principal angles $\theta_{k,\ell}^{(1)},\ldots,\theta_{k,\ell
}^{(d_k\bigvee d_\ell\}}$ between two subspaces $S_k$ and $S_\ell$ of
dimensions $d_k$ and $d_\ell$ are recursively defined by
%
%e2.3 #&#
\[
\label{eqel1} \cos \bigl(\theta^{(i)}_{k\ell} \bigr)=
\max_{\vct{y}\in S_k}\max_{\vct{z} \in
S_\ell} \frac{\vct{y}^T\vct{z}}{\twonorm{\vct{y}}\twonorm{\vct
{z}}} := \frac{\vct{y}_i^T\vct{z}_i}{\twonorm{\vct{y}_i}\twonorm{\vct
{z}_i}},
\]
with the orthogonality constraints $\vct{y}^T\vct{y}_j = 0$,
$\vct{z}^T\vct{z}_j=0$, $j=1,\ldots,i-1$.
\end{definition}
Alternatively, if the columns of $\mtx{U}^{(k)}$ and
$\mtx{U}^{(\ell)}$ are orthobases, then the cosine of the principal
angles are the singular values of
${\mtx{U}^{(k)}}^T\mtx{U}^{(\ell)}$. We write the smallest principal
angle as $\theta_{k\ell} = \theta_{k\ell}^{(1)}$ so that
$\cos(\theta_{k\ell})$ is the largest singular value of
${\mtx{U}^{(k)}}^T\mtx{U}^{(\ell)}$.

%de2.7 #&#
\begin{definition}
The affinity between two subspaces is defined by
%
%e2.4 #&#
\[
\operatorname{{aff}}(S_k,S_\ell)=\sqrt{
\cos^2\theta_{k\ell
}^{(1)}+\cdots+\cos^2
\theta_{k\ell}^{(d_k\bigvee d_\ell)}}.
\]
\end{definition}
In case the distribution of the points are uniform on their
corresponding subspaces, the Geometric Condition (\ref{GeomCond}) may
be reduced to a simple statement about the affinity. This is the
subject of the next theorem.
%
%th2.8 #&#
\begin{theorem}
\label{th2}
Suppose $N_\ell=\rho_\ell d_\ell+1$ points are chosen on each
subspace ${S_\ell}$ at random, $1 \le\ell\le L$. Then as long as
% Suppose the points
% on subspaces $\{S_\ell\}_{\ell=1}^L$ are chosen uniformly at random
%
%e2.5 #&#
\begin{eqnarray}
\label{affcond}
&&\mathop{\max}_{k   :   k \neq\ell} 4\sqrt{2} \bigl(\log
\bigl[N_\ell(N_k+1) \bigr] +\log L+t \bigr)
\frac{\operatorname{aff}(S_k,S_\ell)}{\sqrt{d_k}}
\nonumber
\\[-8pt]
\\[-8pt]
\nonumber
&&\qquad<c(\rho_\ell)\sqrt{\log \rho_\ell}
\qquad\mbox{for each } \ell,
\end{eqnarray}
the subspace detection property holds with probability at least
\[
1-\sum_{\ell=1}^LN_\ell
e^{-\sqrt{d_\ell}\sqrt{N_\ell-1}}-\frac{1}{L^2}\sum_{k\neq\ell}
\frac
{4e^{-2t}}{(N_k+1)N_\ell}.\vadjust{\goodbreak}
\]
Hence, ignoring log factors, subspace clustering is possible if the
affinity between the subspaces is less than about the square root of
the dimension of these subspaces.
\end{theorem}
To derive useful results, assume for simplicity that we have $L$
subspaces of the same dimension $d$ and $\rho d + 1$ points per
subspace so that $N = L(\rho d + 1)$. Then perfect clustering occurs
with probability at least $1-Ne^{-\sqrt{\rho}d}-\frac{2}{(\rho d)(\rho
d+1)}e^{-2t}$ if
%
%e2.6 #&#
\begin{equation}
\label{simplifiedcond} \frac{\operatorname{aff}(S_k,S_\ell)}{\sqrt{d}} < \frac{c(\rho)\sqrt {\log\rho}}{4\sqrt{2}(2\log N+t)}.
\end{equation}
Our notion of affinity matches our basic intuition. To be sure, if the
subspaces are
too close to each other (in terms of our defined notion of affinity),
subspace clustering is hard. Having said this, our result has an
element of surprise. Indeed, the affinity can at most be $\sqrt{d}$
($\sqrt{d_k}$ in general) and, therefore, our result essentially
states that if the affinity is less than $c \sqrt{d}$, then SSC
works. Now this allows for subspaces to intersect and, yet, SSC still
provably clusters all the data points correctly!
%To understand the implications of the above condition consider the
%case that the subspaces are all of dimension $d$ and are randomly
%oriented. Furthermore assume that the number of subspaces $L=
%therefore the right-hand side of (\ref{condition}) is $\mathcal{O}(
%rigorously, the update is that I think I've found a theorem for when
%both $d$ and $n$ large with $\frac{d}{n}=\gamma$ this is true i.e. $
%sure}. This implies that condition (\ref{condition}) reduces to $d<c
%are linear in terms of the ambient dimension, $d=cn$ with $c$ small
%enough, we have precise segmentation.

%To discuss other aspects of this result, assume as before that all
%subspaces have the same dimension $d$. When $d$ is small and the total
%number of subspaces is $\mathcal{O}({n}/{d})$, the problem is
%inherently ``hard'', because it involves clustering all the points
%into many small subgroups. This is reflected by the low probability of
%success in Theorem~\ref{th2}. Of course if one increases the number of
%points chosen from each subspace, the problem should intuitively
%become ``easier''. The probability associated with (
%allows for such a trend. In other words, when $d_\ell$ is small, one
%can increase the probability of success by increasing
%$N_\ell$. Furthermore, with $0<\beta\le1$, the condition can be
%changed to
%which holds with probability at least $1-\sum_{\ell=1}^LN_\ell
%e^{-d_\ell^\beta{(N_\ell-1)}^{(1-\beta)}}-\frac{1}{L^2}\sum_{k\neq\ell}
%if $d_\ell$ is a small fraction of the right-hand side in
%(\ref{affcond}), the subspace detection property holds with much
%higher probability, as expected.

To discuss other aspects of this result, assume as before that all
subspaces have the same dimension $d$. When $d$ is small and the total
number of subspaces is $\mathcal{O}({n}/{d})$, the problem is
inherently hard because it involves clustering all the points into
many small subgroups. This is reflected by the low probability of
success in Theorem~\ref{th2}. Of course, if one increases the number of
points chosen from each subspace, the problem should intuitively
become easier. The probability associated with (\ref{simplifiedcond})
allows for such a trend. In other words, when $d$ is small, one can
increase the probability of success by increasing $\rho$. Introducing
a parameter $0< \beta\le1$, the condition can be modified to
%
%e2.7 #&#
\begin{equation}
\frac{\operatorname{aff}(S_k,S_\ell)}{\sqrt{d}} < \frac{c(\rho)\sqrt {\beta\log\rho}}{4 (2\log N+t)},
\end{equation}
which holds with probability at least
$1-Ne^{-\rho^{(1-\beta)}d}-\frac{2}{(\rho d)(\rho
d+1)}e^{-2t}$. The more general condition (\ref{affcond})
and the corresponding probability can also be modified in a similar
manner.

%s2.2 #&#
\subsection{Segmentation with outliers}

\begin{figure}

\includegraphics{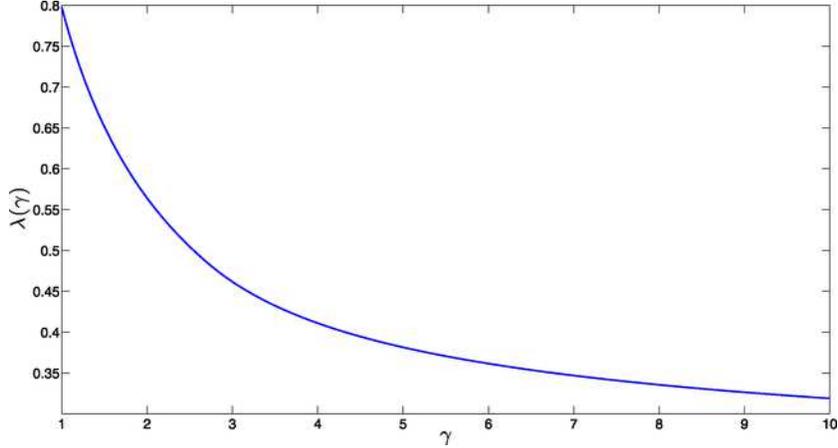}

\caption{Plot of the threshold function (\protect\ref{equthresh}).}\label{figthreshold}
\end{figure}

To see how Algorithm~\ref{algoutlier} works in the presence of
outliers, we begin by introducing a proper threshold function and
define
%
%e2.8 #&#
\begin{equation}
\label{equthresh} \lambda(\gamma) = %
\cases{ \displaystyle\sqrt{
\frac{2}{\pi}}\frac{1}{\sqrt{\gamma}},&\quad $1\le\gamma\le e, $\vspace*{2pt}
\cr
\displaystyle\sqrt{ \frac{2}{\pi
e}}\frac{1}{\sqrt{\log\gamma}}, &\quad $\gamma\ge e,$ } %
\end{equation}
%
%Notice that this function only depends on the $\rho$, i.e. it only
%depends on the ratio between the dimension and the number of points
%under consideration.
shown in Figure~\ref{figthreshold}. The theorem below justifies the
claims made in the introduction.
%
%f5 #&#

% \begin{theorem}
% \label{th3}
% Suppose the outlier points are chosen uniformly at random and set
% $\frac{1}{\gamma}=\frac{N-1}{n}$.
% \begin{itemize}
% \item[(a)] If
% \begin{equation}
% \label{OutlierCondGeom}
% \underset{\ell,i}{\max}\frac{1}{r(\mathcal{P}(\mtx{X}^{(
% %=\begin{cases} \frac{1}{\sqrt{e}}\sqrt{\frac{2}{\pi}}(1-t)\frac{1}{{(
%}} & \mbox{if } \text{ }\text{ }\text{ }\text{ }\frac{N-1}{n}\ge e
% \end{equation}
% then all the outliers are identified correctly with probability at
% least $1-N_0e^{-C_1t^2\frac{n}{\log(2(N-1))}}$ for some positive
% numerical constant $C_1$.
% \item[(b)] Furthermore, assume the points on the subspaces are chosen
% uniformly at random. If
% \begin{equation}
% \label{OutlierCond}
% \underset{\ell}{\max}\frac{\sqrt{2d_\ell}}{c(\rho_\ell)\sqrt{\log
%}} & \mbox{if } \text{ }\text{ }\text{ }\text{ }\frac{N-1}{n}\ge e
% \end{equation}
% then with probability at
% least $1-\sum_{\ell=1}^{L}N_\ell e^{-\sqrt{d_\ell}\sqrt{(N_\ell-1)}}$,
% no `real' data point is wrongfully detected as an outlier.
% \end{itemize}
% \end{theorem}
%
%th2.9 #&#
\begin{theorem}
\label{th3}
Suppose the outlier points are chosen uniformly at random and set
$\gamma=\frac{N-1}{n}$, then using the threshold value
$(1-t)\frac{\lambda(\gamma)}{\sqrt{e}}\sqrt{n}$, all outliers are
identified correctly with probability at least
$1-N_0e^{-C_1t^2{n}/{\log N}}$ for some positive numerical
constant $C_1$. Furthermore, we have the following guarantees in the
deterministic and semi-random models:
\begin{longlist}[(a)]
\item[(a)] If in the deterministic model,
%
%e2.9 #&#
\begin{equation}
\label{OutlierCondGeom} \mathop{\max}_{\ell,i}\frac{1}{r(\mathcal{P}(\mtx{X}^{(\ell
)}_{-i}))}<(1-t)
\frac{\lambda(\gamma)}{\sqrt{e}}\sqrt{n},
\end{equation}
then no ``real'' data point is wrongfully detected as an outlier.
\item[(b)] If in the semi-random model,
%
%e2.10 #&#
\begin{equation}
\label{OutlierCond} \mathop{\max}_{\ell}\frac{\sqrt{2d_\ell}}{c(\rho_\ell)\sqrt{\log\rho_\ell}}<(1-t)
\frac{\lambda(\gamma)}{\sqrt{e}}\sqrt{n},
\end{equation}
then with probability at
least $1-\sum_{\ell=1}^{L}N_\ell e^{-\sqrt{d_\ell}\sqrt{(N_\ell-1)}}$,
no ``real'' data point is wrongfully detected as an outlier.
\end{longlist}

\end{theorem}
The threshold in the right-hand side of (\ref{OutlierCondGeom}) and
(\ref{OutlierCond}) is essentially $\sqrt{n}$ multiplied by a factor
which depends only on the ratio of the number of points and the
dimension of the ambient space.

As in the situation with no outliers, when $d_\ell$ is small we need
to increase $N_\ell$ to get a result holding with high
probability. Again this is expected because when $d_\ell$ is small, we
need to be able to separate the outliers from many small clusters
which is inherently a hard problem for small values of~$N_\ell$.

%To see what the implications of condition (\ref{OutlierCond}) are in
%terms of the number of outliers, assume without loss of generality
%that the maximum value of $\sqrt{d_\ell/\log(\frac{N_\ell-1}{d_
%generality assume that $\frac{N-1}{n}\ge e$ (the other case is
%similar), then the condition (\ref{OutlierCond}) implies that $N_0<n{(
%Notice that ${(\frac{N-1}{n})}^{\frac{1}{2(1+\log\frac{N-1}{n})}}\le
%Now suppose $d_1=\mathcal{O}(\sqrt{n})$, this implies that we can
%detect $\mathcal{O}(n{(\frac{N_1-1}{d_1})}^{\mathcal{O}(\sqrt{n})})$
%outliers correctly, which is exponential in terms of $n$, i.e. we can
%detect exponentially many outliers with high probability.

The careful reader will notice a factor $\sqrt{e}$ discrepancy between
the threshold $\lambda(\gamma)\sqrt{n}$ presented in Algorithm
\ref{algoutlier} and what is proven in (\ref{OutlierCondGeom}) and
(\ref{OutlierCond}). We believe that this is a result of our
analysis\footnote{More specifically, from switching from the mean
width to a volumetric argument by means of Urysohn's inequality.}
and we conjecture that (\ref{OutlierCondGeom}) and (\ref{OutlierCond})
hold without the factor $\sqrt{e}$ in the denominator. Our simulations
in Section~\ref{numeric} support this conjecture.

%s3 #&#
\section{Discussion and comparison with other work}
\label{comparison}

It is time to compare our results with a couple of previous important
theoretical advances. To introduce these earlier works, we first need
some definitions.
%
%de3.1 #&#
\begin{definition}
The subspaces $\{S_\ell\}_{\ell=1}^L$ are said to be independent if
and only if $\sum_{\ell} \operatorname{\rm dim}(S_\ell) = \operatorname
{\rm
dim}(\oplus_{\ell} S_\ell)$, where $\oplus$ is the direct sum.
\end{definition}
For instance, three lines in $\mathbb{R}^2$ cannot be independent.
%
%de3.2 #&#
\begin{definition}
The subspaces $\{S_\ell\}_{\ell=1}^L$ are said to be disjoint if and
only if for all pairs $k \neq\ell$, $S_k\cap S_\ell
=\{\vct{0}\}$.
\end{definition}

% \begin{definition}
% For a matrix $\mtx{A} \in\mathbb{R}^{n\times N}$, let $\mathbb{W}_d(
%have dimension $n\times d$.
% \end{definition}
%
%de3.3 #&#
\begin{definition}
The geodesic distance between two subspaces $S_i$ and $S_j$ of
dimension $d$, denoted by $\operatorname{{dist}}(S_i,S_j)$, is defined
by
%
%e3.1 #&#
\[
\operatorname{{dist}}(S_k,S_\ell)=\sqrt{
\sum_{i=1}^{d_k\bigvee
d_\ell} \bigl(\theta_{k\ell}^{(i)}
\bigr)^2}.
\]
\end{definition}
%
%s3.1 #&#
\subsection{Segmentation without outliers}
In \cite{Elhamifar09}, Elhamifar and Vidal show that the subspace
detection property holds as long as the subspaces are independent. In~\cite{Elhamifar10},
the same authors show that under less restrictive
conditions the $\ell_1$ subspace detection property still holds. %In
%order to introduce their precise statement, consider a collection of
%data points drawn from $L$ subspaces $S_1,\ldots,S_L$ of corresponding
%dimensions $d_1,\ldots,d_\ell$. Furthermore, we assume that the $N_
Formally, they show that if
%
%e3.2 #&#
\begin{equation}
\label{Vidalcond} \frac{1}{\sqrt{d_\ell}}  \mathop{
\max}_{\mtx{Y}\in\mathbb{W}_{d_\ell}(\mtx{X^{(\ell)}})}\sigma_{\min}(\mtx {Y})>\mathop{
\max}_{k: k\neq\ell}\cos \bigl(\theta_{k\ell}^{(1)} \bigr)\qquad
\mbox{for all } \ell=1, \ldots, L,
\end{equation}
then the subspace detection property holds. In the above formulation,
$\sigma_{\min}(\mtx{Y})$ denotes the smallest singular value of $\mtx{Y}$
and $\mathbb{W}_d(\mtx{X}^{(\ell)})$ denotes the set of all full rank
sub-matrices of $\mtx{X}^{(\ell)}$ of size $n\times d_\ell$. The
interesting part of the above condition is the appearance of the
principal angle on the right-hand side. However, the left-hand side is
not particularly insightful (i.e., it does not tell us anything about\vadjust{\goodbreak}
the important parameters involved in the subspace clustering problem,
such as dimensions, number of subspaces and so on) and it is in fact
NP-hard to even calculate it.
\begin{itemize}
\item\textit{Deterministic model.} This paper also introduces a
sufficient condition \eqref{GeomCond} under which the subspace
detection property holds in the fully deterministic setting; compare
Theorem~\ref{th1}. This sufficient condition is much less
restrictive as any configuration obeying (\ref{Vidalcond}) also
obeys \eqref{GeomCond}. More precisely,
$\mu(\mathcal{X}_\ell)\le{\max}_{k: k\neq
\ell}\cos(\theta_{k\ell}^{(1)})$ and
$\frac{1}{\sqrt{d_\ell}}   {\max}_{\mtx{Y}\in\mathbb{W}_{d_\ell}(\mtx{X^{(\ell)}})}\sigma_{\min}(\mtx{Y})\le\break
\min_ir(\mathcal{P}_{-i}^\ell)$.\footnote{The
latter follows from
$\max_i\frac{1}{r(\mathcal{P}_{-i}^\ell)}\le{\min}_{\mtx{Y}\in\mathbb
{W}_{d_\ell}(\mtx{X^{(\ell)}})}\frac{\sqrt{d_\ell}}{\sigma_{\min
}(\mtx{Y})}$
which is a simple consequence of Lemma~\ref{step1}.} As for
\eqref{Vidalcond}, checking that \eqref{GeomCond} holds is also
NP-hard in general. However, to prove that the \sdp\ holds, it is
sufficient to check a slightly less restrictive condition than
\eqref{GeomCond}; this is tractable, see Lemma~\ref{dualcertificate}. %The reason is that computing the
%inradius of a polytope is a linear program, see \cite[Chapter
%4]{BoydBook}.

\item\textit{Semi-random model.} Assume that all subspaces are of the
same dimension $d$ and that there are $\rho d+1$ points on each
subspace. Since the columns of $\mtx{Y}$ have unit norm, it is easy
to see that the left-hand side of (\ref{Vidalcond}) is strictly less
than $1/\sqrt{d}$. Thus, \eqref{Vidalcond} at best restricts the
range for perfect subspace recovery to $\cos
\theta_{k\ell}^{(1)} <c\frac{1}{\sqrt{d}}$ [by looking at
\eqref{Vidalcond}, it is not entirely clear that this would even be
achievable]. In comparison, Theorem~\ref{th2} (excluding some
logarithmic factors for ease of presentation) requires
%
%e3.3 #&#
\begin{eqnarray}
\label{simplecond} \operatorname{aff}(S_k,S_\ell)&=&\sqrt
{\cos^2 \bigl(\theta_{k\ell}^{(1)} \bigr)+
\cos^2 \bigl(\theta_{k\ell}^{(2)} \bigr)+\cdots+
\cos^2 \bigl(\theta_{k\ell}^{(d)} \bigr)}
\nonumber
\\[-8pt]
\\[-8pt]
\nonumber
&<&c\sqrt{
\log( \rho )}\sqrt{d}.
\end{eqnarray}
The left-hand side can be much smaller than $\sqrt{d}\cos
\theta_{k\ell}^{(1)}$ and is, therefore, less restrictive.

To be more specific, assume that in the model described above we have
two subspaces with an intersection of dimension $s$. Because the two
subspaces intersect, the condition given by Elhamifar and Vidal
becomes $1<\frac{1}{\sqrt{d}}$, which cannot hold. In comparison, our
condition \eqref{simplecond} simplifies to
\[
\cos^2 \bigl(\theta_{k\ell}^{(s+1)} \bigr)+\cdots+
\cos^2 \bigl(\theta_{k\ell}^{(d)} \bigr)<c\log (
\rho)d-s,
\]
which holds as long as $s$ is not too large and/or a fraction of the
angles are not too small. From an application standpoint, this is
important because it explains why SSC can often succeed even when the
subspaces are not
disjoint. %To give more detail, this condition holds for example when
%the two intersecting subspaces are orthogonal to each other
%in all directions other than the intersecting line (i.e. if
%$\cos(\theta^{(2)})=\ldots=\cos(\theta^{(d)})=0$). Furthermore,
%it also allows for a tradeoff between the number of points
%on each subspace and the principal angles; excluding the
%first $s$ principal angles.
%
\item\textit{Fully random model.} As before, assume for simplicity
that all subspaces are of the same dimension $d$ and that there are
$\rho d+1$ points on each subspace. We have seen that
\eqref{Vidalcond}\vadjust{\goodbreak} imposes $\cos\theta_{k\ell}^{(1)} < c\frac{1}{\sqrt {d}}$. It can
be shown that in the fully random setting,\footnote{One can see this
by noticing that the square of this parameter is the largest root
of a multivariate beta distribution. The asymptotic value of this
root can be calculated, for example, see \cite{Johnstone08}.} $\cos
\theta_{k\ell}^{(1)} \approx c\sqrt{\frac{d}{n}}$. Therefore, \eqref{Vidalcond}
would put a restriction of the form
\[
d<c\sqrt{n}.
\]
In comparison,
Theorem~\ref{th0} requires
\[
d<c_1\frac{\log\rho}{ \log N}n,
\]
which allows for the dimension of the subspaces to be almost linear in
the ambient dimension.
\end{itemize}

Such improvements come from a geometric insight: it becomes
apparent that the SSC algorithm succeeds if the actual subspace
points (primal directions) have small inner products with the dual
directions on another subspace. This is in contrast with Elhamifar
and Vidal's condition which requires that the inner products between
\textit{any} direction on one subspace and \textit{any} direction on
another be small. Further geometric explanations are given in
Section~\ref{geometricview}.
% This should be contrasted with Elhamifar and Vidals
% condition which require that the dot product between \textit{any}
% direction on one subspace and \textit{any} direction on the other to be
% small. The latter is of course much more restrictive. We also provide
% a geometric explanation of this in the concluding paragraph of section
%~\ref{geometricview}.

%s3.2 #&#
\subsection{Segmentation with outliers}

To the best of our knowledge, there is only one other theoretical
result regarding outlier detection. In \cite{Lerman10}, Lerman
and Zhang study the effectiveness of recovering subspaces in the
presence of outliers by some sort of $\ell_p$ minimization for
different values of $0<p<\infty$. They address simultaneous recovery
of all $L$ subspaces by minimizing the functional
%
%e3.4 #&#
\begin{equation}
\label{energy} e_{\ell_p}(\mathcal{X},S_1,
\ldots,S_L)=\sum_{\vct{x}\in\mathcal
{X}}\mathop{
\min}_{1\le\ell\le L}{ \bigl(\operatorname{dist}(\vct{x},S_\ell)
\bigr)}^p.
\end{equation}
Here, $S_1,\ldots,S_L$ are the optimization variables and
$\mathcal{X}$ is our data set. This is not a convex optimization for
any $p>0$, since the feasible set is the Grassmannian.

% In the semi-random model, Lerman et. al. show that under the
% assumptions stated in Theorem~\ref{th0outlier}, with $0<p\le1$ and
% $\lambda_1(d)$ a constant depending on $d$, the subspaces
% $S_1,\ldots,S_L$ minimize (with large probability) the energy
% (\ref{energy}) among all $d$-subspaces in $\mathbb{R}^n$ if
% \begin{eqnarray}
% \label{Lermancond}
% N_0<\frac{\rho d}{2}\min\bigg(\lambda_1(d),\frac{1}{4Ld^{\frac{3}{2}}}
% \end{eqnarray}
% It is easy to see that the right-hand side of (\ref{Lermancond}) is
% upperbounded by $\rho d$, i.e. the typical number of points on each
% subspace. Notice that our analogous result in Theorem~\ref{th0},
% allows for a much larger number of outliers. In fact, the number of
% outliers can sometimes even be much larger than the total number of
% data points on all the other subspaces combined. Our proposed
% algorithm also has the added benefit that it is convex and,
% therefore, practical. Having said this, it is worth mentioning that an
% interesting byproduct of the result from Lerman et. al. is that the
% energy minimization can perform perfect subspace recovery when no
% outliers are present. In fact, they even extend this to the case when
% the subspace points are noisy.

In the semi-random model, the result of Lerman and Zhang states that
under the assumptions stated in Theorem~\ref{th0outlier}, with
$0<p\le1$ and $\tau_0$ a constant,\footnote{The result of
\cite{Lerman10} is a bit more general in that the points on each
subspace can be sampled from a single distribution obeying certain
regularity conditions, other than the uniform measure. In this case,
$\tau_0$ depends on this distribution as well.} the subspaces
$S_1,\ldots,S_L$ minimize (with large probability) the energy
\eqref{energy} among all $d$-dimensional subspaces in $\mathbb{R}^n$
if
%
%e3.5 #&#
\begin{equation}
\label{Lermancond} N_0<\tau_0\rho d\min \Bigl(1,
\mathop{\min}_{k\neq\ell}{\operatorname{ dist}(S_k,S_\ell)^p}/{2^p}
\Bigr).
\end{equation}
It is easy to see that the right-hand side of (\ref{Lermancond}) is
upperbounded by $\rho d$, that is, the typical number of points on each
subspace. Notice that our analogous result in Theorem~\ref{th0} allows
for a much larger number of outliers. In fact, the number of outliers
can sometimes even be much larger than the total number of data points
on all subspaces combined. Our proposed algorithm also has
the added benefit that it is convex and, therefore, practical. Having
said this, it is worth mentioning that the results in \cite{Lerman10}
hold for a more general outlier model.
% Furthermore, in certain cases the condition above can be some what
% weakened by iterative use of this method as stated in
% \cite{Lermaniter11}.
Also, an interesting byproduct of the result from Lerman and Zhang is
that the energy minimization can perform perfect subspace recovery
when no outliers are present. In fact, they even extend this to the
case when the subspace points are noisy.

Finally, while this manuscript was in preparation, Liu Guangcan
brought to our attention a new paper \cite{Liu11}, which also
addresses outlier detection. However, the suggested scheme limits the
number of outliers to $N_0<n-\sum_{\ell=1}^Ld_\ell$. That is, when the
total dimension of the subspaces ($\sum_{\ell=1}^Ld_\ell$) exceeds the
ambient dimension $n$, outlier detection is not possible based on the
suggested scheme. In contrast, our results guarantee perfect outlier
detection even when the number of outliers far exceeds the number of
data points.

%s4 #&#
\section{Geometric perspective on the separation condition}
\label{Geompers}

The goal of this section is twofold. One aim is to provide a geometric
understanding of the \sdp\ and of the sufficient condition presented in
Section~\ref{WOth}. Another is to introduce concepts such as $\mathcal{K}$-norms
and polar sets, which will play a crucial role in our analysis.

%s4.1 #&#
\subsection{Linear programming theory}

We are interested in finding the support of the
%f6 #&#
\begin{figure}

\includegraphics{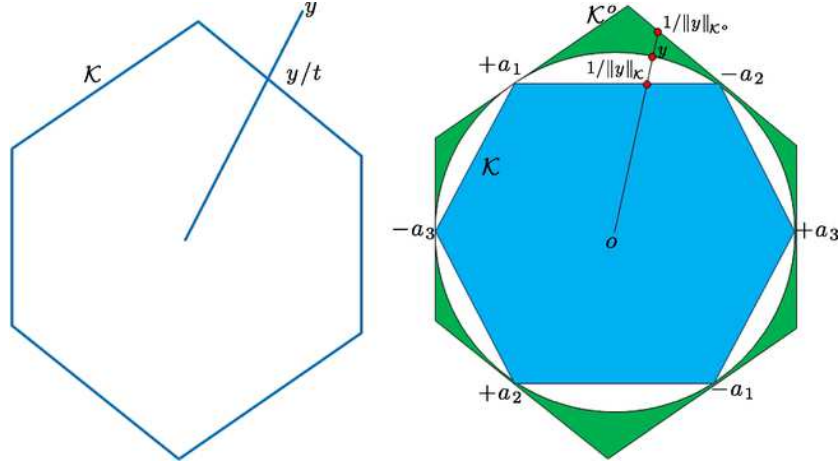}

\caption{Illustration of Definitions \protect\ref{defKnorm} and
\protect\ref{defpolar}. \textup{(a)} Norm with respect to a polytope $\mathcal{K}$. \textup{(b)}
Polytope $\mathcal{K}$ and its polar $\mathcal{K}^o$.}\label{fignormandpolar}%{fig:norm}{fig:dual}
\end{figure}
optimal solution to
%
%e4.1 #&#
\begin{equation}
\label{eqprimal} \min_{\vct{x} \in\R^N} \|\vct{x}\|_{\ell_1} \qquad\mbox{subject
to } \mtx{A}\vct{x} = \vct{y},
\end{equation}
where both $\vct{y}$ and the columns of $\mtx{A}$ have unit
norm. The dual takes the form
%
%e4.2 #&#
\begin{equation}
\label{dualeq} \max_{\vct{z} \in\R^n} \langle\vct{y} , \vct{z} \rangle
\qquad\mbox{subject to } \bigl\|\mtx{A}^T\vct{z}\bigr\|_{\ell_\infty} \le1.
\end{equation}
Since strong duality always holds in linear programming, the optimal
values of~\eqref{eqprimal} and \eqref{dualeq} are equal. We now
introduce some notation to express the dual program differently.
%
%de4.11 #&#
\begin{definition}
\label{defKnorm}
The norm of a vector $\vct{y}$ with respect to a symmetric convex
body is defined as
%
%e4.3 #&#
\begin{equation}
\|\vct{y}\|_\mathcal{K}=\inf \{t>0 \dvtx \vct{y}/{t}\in\mathcal {K} \}.
\end{equation}
\end{definition}

This norm is shown in Figure~\ref{fignormandpolar}(a).\vadjust{\goodbreak}
%
%de4.2 #&#
\begin{definition}
\label{defpolar}
The polar set $\mathcal{K}^o$ of $\mathcal{K} \subset\R^n$ is defined as
%
%e4.4 #&#
\begin{equation}
\mathcal{K}^o= \bigl\{\vct{y}\in\mathbb{R}^n\dvtx
\langle\vct{x} , \vct{y} \rangle\le1 \mbox{ for all } \vct{x}\in\mathcal{K} \bigr
\}.
\end{equation}
\end{definition}

Set $\mathcal{K}^o =  \{\vct{z}\dvtx  \infnorm{\mtx{A}^T\vct{z}}\le
1 \}$ so that our dual problem \eqref{dualeq} is of the form
%
%e4.5 #&#
\begin{equation}
\max_{\vct{z} \in\R^n} \langle\vct{y} , {z} \rangle \qquad\mbox{subject to } \vct{z}\in
\mathcal{K}^o.
\end{equation}

It then follows from the definitions above that the optimal value of
\eqref{eqprimal} is given by $\|\vct{y}\|_{\mathcal{K}}$, where
$\mathcal{K} =
\operatorname{conv} (\pm\vct{a}_1,\ldots,\pm\vct{a}_N )$; that is
to say, the minimum value of the $\ell_1$ norm is the norm of
$\vct{y}$ with respect to the symmetrized convex hull of the
columns of $\mtx{A}$. In other words, this perspective asserts that
support detection in an $\ell_1$ minimization problem is equivalent to
finding the face of the polytope $\mathcal{K}$ that passes through the ray
$\vec{y}=\{t\vct{y}, t \ge0\}$; the extreme points of this face
reveal those indices with a nonzero entry. We will refer to the face
passing through the ray $\vec{y}$ as the face closest to
$\vct{y}$. Figure~\ref{fignormandpolar}(b) illustrates some of these
concepts.

%
%f7 #&#
\begin{figure}

\includegraphics{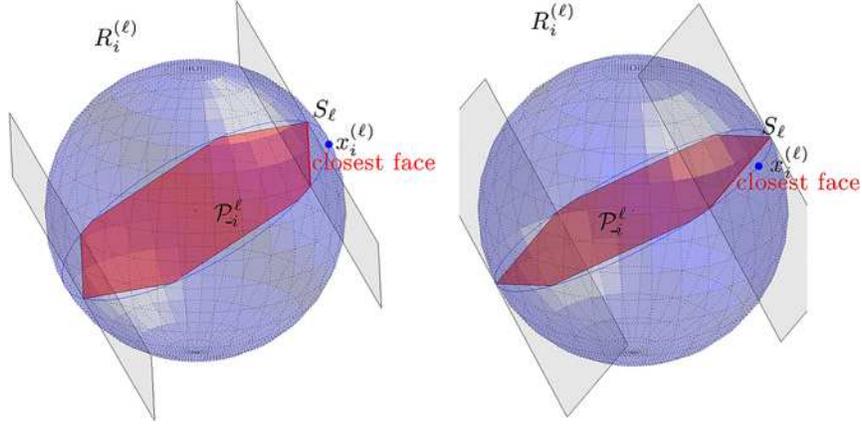}

\caption{Illustration of $\ell_1$ minimization when the subspace
detection property holds. Same object seen from different
angles.}\label{figclosestface}
\end{figure}

%f8 #&#
\begin{figure}[b]

\includegraphics{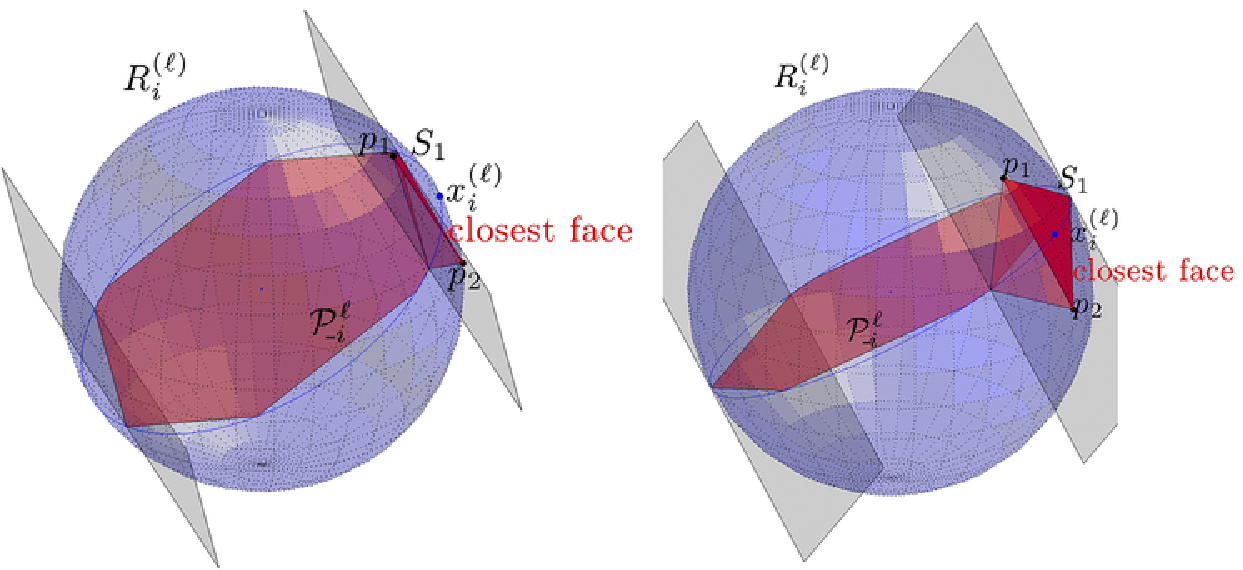}

\caption{Illustration of $\ell_1$ minimization when the subspace
detection property fails. Same object seen from different angles.}\label{fig2points}
\end{figure}

%f9 #&#
\begin{figure}

\includegraphics{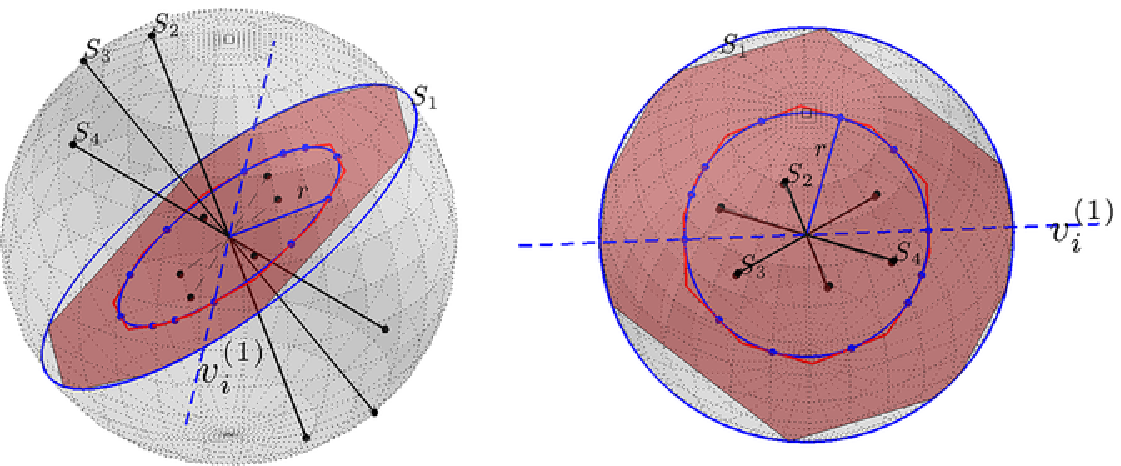}

\caption{Geometric view of \protect\eqref{GeomCond}. The right figure is seen
from a direction orthogonal to~$S_1$.}\label{figIntuition}
\end{figure}

%s4.2 #&#
\subsection{A geometric view of the subspace detection
property}
\label{geometricview}
We have seen that the subspace detection property holds if for each point
$\vct{x}_i$, the closest face to $\vct{x}_i$ resides in the same
subspace. To establish a geometric characterization, consider an
arbitrary point, for instance, $\vct{x}_i^{(\ell)} \in S_\ell$ as in
Figure~\ref{figclosestface}. Now construct the symmetrized convex
hull of all the other points in $S_\ell$ indicated by
$\mathcal{P}_{-i}^\ell$ in the figure. Consider the face of
$\mathcal{P}_{-i}^\ell$ that is closest to $\vct{x}_i^{(\ell)}$; this
face is shown in Figure~\ref{figclosestface} by the line segment in
red. Also, consider the plane passing through this segment and
orthogonal to $S_\ell$ along with its reflection about the origin;
this is shown in Figure~\ref{figclosestface}\vadjust{\goodbreak} by the light grey
planes. Set $R_i^{(\ell)}$ to be the region of space restricted
between these two planes. Intuitively, if no two points on the other
subspaces lie outside of $R_i^{(\ell)}$, then the face chosen by the
algorithm is as in the figure and lies in $S_\ell$.

To illustrate this point further, suppose there are two points not in
$S_\ell$ lying outside of the region $R_i^{(\ell)}$ as in
Figure~\ref{fig2points}. In this case, the closest face does not lie
in $S_\ell$ as can be seen in the figure. Therefore, one could
intuitively argue that a sufficient condition for the closest face to
lie in $S_\ell$ is that the projections onto $S_\ell$ of the points
from all the other subspaces do not lie outside of regions\vadjust{\goodbreak}
$R_i^{(\ell)}$ for all points $\vct{x}_i^{(\ell)}$ in subspace
$S_\ell$. This condition is closely related to the sufficient
condition stated in Theorem~\ref{th1}. More precisely, the dual
directions $\vct{v}_i^{(\ell)}$ approximate the normal directions to
the restricting planes $R_i^{(\ell)}$, and
$\min_{i}r(\mathcal{P}_{-i}^\ell)$ the distance of these planes from
the origin.

Finally, to understand the sufficient condition of Theorem~\ref{th1},
we will use Figure~\ref{figIntuition}. We focus on a single subspace,
say, $S_1$. As previously stated, a sufficient condition is to have all
points not in $S_1$ to have small coherence with the dual directions
of the points in $S_1$. The dual directions are depicted in
Figure~\ref{figIntuition} (blue dots). One such dual direction line
is shown as the dashed blue line in the figure. The points that have
low coherence with the dual directions are the points whose projection
onto subspace $S_1$ lie inside the red polytope. As can be seen, this
polytope approximates the intersection of regions $R_i^{(1)}$
($\bigcap_{i=1}^{N_1} R_i^{(1)}$) and subspace~$S_1$. This helps in
understanding the difference between the condition imposed by
Elhamifar and Vidal and our condition; in this setting, their
condition essentially states that the projection of the points on all
other subspaces onto subspace $S_1$ must lie inside the blue
circle. By looking at Figure~\ref{figIntuition}, one might draw the
conclusion that these conditions are very similar, that is, the red
polytope and the blue ball restrict almost the same region. This is
not the case, because as the dimension of the subspace $S_1$ increases
most of the volume of the red polytope will be concentrated around its
vertices and the ball will only occupy a very small fraction of the
total volume of the polytope.

%s5 #&#
\section{Numerical results}
\label{numeric}

This section proposes numerical experiments on synthesized data to
further our understanding of the behavior/limitations of SSC, of our
analysis and of our proposed outlier detection scheme. In this
numerical study we restrict ourselves to understanding the effect of
noise on the spectral gap and the estimation of the number of
subspaces. For a more comprehensive\vadjust{\goodbreak} analytical and numerical study of
SSC in the presence of noise, we refer the reader to
\cite{Elhamifar11}. For comparison of SSC with more recent methods on
motion segmentation data, we refer the reader to
\cite{Liu10,Favaro11}. % \MS{\cite{Elhamifar12}, which was submitted a
% few months after this paper became available online, also performs
% additional interesting simulations and comparisons with the state of
% the art.}\MS{I want to make the timing of this paper clear, the
% above is probably not the correct way.} \EJC{We should not cite
% [Elhamifar12], period.}
These papers indicate that SSC has the best
performance on the Hopkins 155 data \cite{Tron07} when corrupted
trajectories are present, and has a performance competitive with the
state of the art when there is no corrupted trajectory. In the spirit
of reproducible research, the Matlab code generating all the plots is
available at \url{http://www.stanford.edu/\textasciitilde mahdisol/Software}.

%s5.1 #&#
\subsection{Segmentation without outliers}

As mentioned in the \hyperref[secintro]{Introduction}, the subspace detection property can
hold even when the dimensions of the subspaces are large in comparison
with the ambient dimension $n$. SSC can also work beyond the region
where the subspace detection property holds because of further
spectral clustering. Section~\ref{metrics} introduces several metrics
to assess performance and Section~\ref{intersect} demonstrates that
the subspace detection property can hold even when the subspaces
intersect. In Section~\ref{affvsnum} we study the performance of SSC
under changes in the affinity between subspaces and the number of
points per subspace. In Section~\ref{dim} we illustrate the effect of
the dimension of the subspaces on the subspace detection property and
the spectral gap. In Section~\ref{noiseModel} we study the effect of
noise on the spectral gap. In the final subsection we study the
capability of SSC in estimating the correct number of subspaces and
compare it with a classical algorithm.

%s5.1.1 #&#
\subsubsection{Error metrics}

\label{metrics}
The four different metrics we use are as follows (see
\cite{Elhamifar10} for simulations using similar metrics):
\begin{itemize}
\item\textit{Feature detection error.} For each point $\vct{x}_i$,
partition the
optimal solution of SSC as
\[
\vct{z}_i=\mct{\Gamma} %
\left[\matrix{\vct{z}_{i1}
\vspace*{2pt}
\cr
\vct{z}_{i2}\vspace*{2pt}
\cr
\vdots\vspace*{2pt}
\cr
\vct{z}_{iL} }\right] %
.
\]
In this representation, $\mct{\Gamma}$ is our unknown permutation
matrix and $\vct{z}_{i1}, \vct{z}_{i2},\break \ldots,\vct{z}_{iL}$ denote the
coefficients corresponding to each of the $L$ subspaces. Using
$N$ as the total number of points, the feature detection error
is
%
%e5.1 #&#
\begin{equation}
\label{eqdetectionerror} \frac{1}{N}\sum_{i=1}^{N}
\biggl(1-\frac{\onenorm{\vct
{z}_{ik_i}}}{\onenorm{\vct{z}_i}} \biggr),
\end{equation}
in which $k_i$ is the subpace $\vct{x}_i$ belongs to. The quantity
between brackets in \eqref{eqdetectionerror} measures how far we
are from choosing all our neighbors in the same subspace; when the
subspace detection property holds, this term is equal to $0$ whereas
it takes on the value $1$ when all the points are chosen from the
other subspaces.\vadjust{\goodbreak}

\item\textit{Clustering error.} Here, we assume knowledge of the number
of subspaces and apply spectral clustering to the affinity matrix
built by the SSC algorithm. After the spectral clustering step, the
clustering error is simply defined as
%
%e5.2 #&#
\begin{equation}
\frac{\#\mbox{ of misclassified points}}{\mbox{total } \#\mbox{ of points}}.
\end{equation}

\item\textit{Error in estimating the number of subspaces.} This is a
0-1 error which takes on the value $0$ if the true number of
subspaces is correctly estimated, and $1$ otherwise.

\item\textit{Smallest nonzero eigenvalue.} We use the $(N-L)+1$th
smallest eigenvalue of the normalized Laplacian\footnote{After
building the symmetrized affinity graph
$\mtx{W}=\abs{\mtx{Z}}+\abs{\mtx{Z}}^T$, we form the normalized
Laplacian $\mtx{L}_N=\mtx{I}-\mtx{D}^{-1/2}\mtx{W}\mtx{D}^{-1/2}$,
where $\mtx{D}$ is a diagonal matrix and $D_{ii}$ is equal to the
sum of the elements in column $\mtx{W}_i$. This form of the
Laplacian works better for spectral clustering as observed in many
applications \cite{Ng02}.} as a numerical check on whether the
subspace detection property holds (when the subspace detection
property holds this value vanishes).
\end{itemize}

%s5.1.2 #&#
\subsubsection{Subspace detection property holds even when the
subspaces intersect}
\label{intersect}

We wish to demonstrate that the subspace detection property holds even
when the subspaces intersect. To this end, we generate two subspaces
of dimension $d=10$ in $\R^{n=200}$ with an intersection of dimension
$s$. We sample one subspace ($S_1$) of dimension $d$ uniformly at
random among all $d$-dimensional subspaces and a subspace of dimension
$s$ [denoted by $S_2^{(1)}$] inside that subspace, again, uniformly at
random. Sample another subspace $S_2^{(2)}$ of dimension $d-s$
uniformly at random and set $S_2=S_2^{(1)}\oplus S_2^{(2)}$.

%f10 #&#
\begin{figure}

\includegraphics{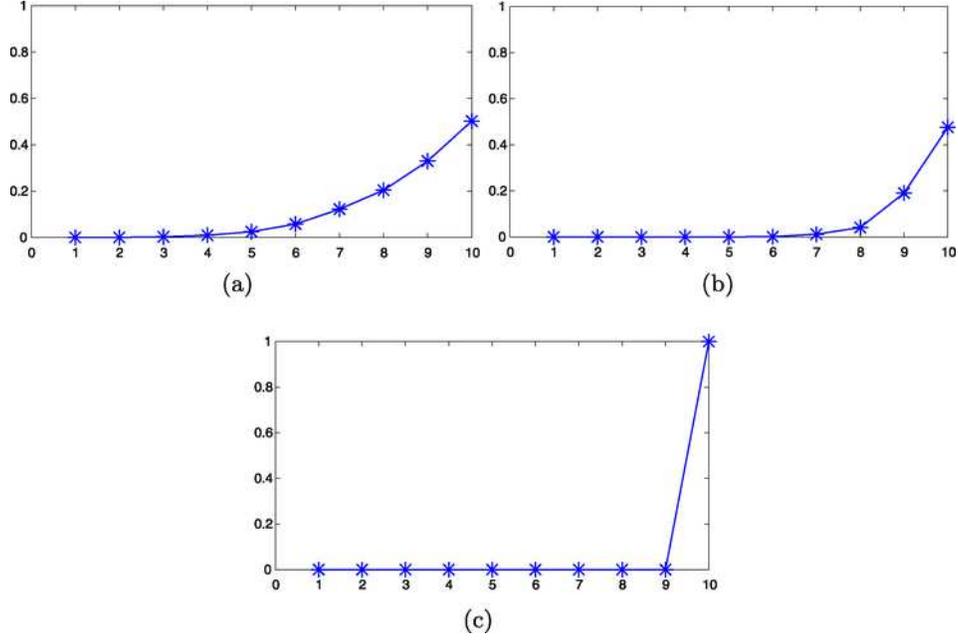}

\caption{Error metrics as a function of the dimension of the
intersection. \textup{(a)} Feature detection error. \textup{(b)} Clustering error. \textup{(c)}
Error in estimating the number of subspaces.}\label{figinter}
\end{figure}

Our experiment selects $N_1=N_2=20d$ points uniformly at random from
each subspace. We generate $20$ instances from this model and report
the average of the first three error criteria over these instances;
see Figure~\ref{figinter}. Here, the subspace detection property
holds up to $s=3$. Also, after the spectral clustering step, SSC has a
vanishing clustering error even when the dimension of the intersection
is as large as $s = 6$.

%s5.1.3 #&#
\subsubsection{Effect of the affinity between subspaces}
\label{affvsnum}

In Section~\ref{semirandomsec} we showed that in the semi-random
model, the success of SSC depends upon the affinity between the
subspaces and upon the density of points per subspace (recovery
becomes harder as the affinity increases and as the density of points
per subspace decreases). We study here this trade-off in greater
detail through experiments on synthetic data.

We generate $3$ subspaces $S_1$, $S_2$ and $S_3$, each of dimension
$d=20$ in $\R^{n=40}$. The choice $n=2d$ makes the problem challenging
since every data point on one subspace can also be expressed as a
linear combination of points on other subspaces. The bases we choose
for $S_1$ and $S_2$ are
%
%e5.3 #&#
\begin{eqnarray}
\mtx{U}^{(1)}= %
\left[\matrix{\mtx{I}_d \vspace*{2pt}
\cr
\mtx{0}_{d\times d} }\right]
,\qquad  \mtx{U}^{(2)}= \left[
\matrix{\mtx{0}_{d\times d} \vspace*{2pt}
\cr
\mtx{I}_d }\right]
,
\end{eqnarray}
whereas for $S_3$,
%
%e5.4 #&#
\begin{equation}
\mtx{U}^{(3)}= %
\left[\matrix{\cos(\theta_1)&0&0&0&
\ldots&0\vspace*{2pt}
\cr
0&\cos(\theta_2)&0&0&\ldots&0\vspace*{2pt}
\cr
0&0&\cos(\theta_3)&0&\ldots&0\vspace*{2pt}
\cr
\vdots&\vdots&
\vdots & \vdots&\ddots&\vdots\vspace*{2pt}
\cr
0&0&0&0&\ldots&\cos(
\theta_d) \vspace*{2pt}
\cr
\sin(\theta_1)&0&0&0&\ldots&0
\vspace*{2pt}
\cr
0& \sin(\theta_2)&0&0&\ldots&0\vspace*{2pt}
\cr
0&0&
\sin( \theta_3)&0&\ldots&0\vspace*{2pt}
\cr
\vdots&\vdots&\vdots&
\vdots& \ddots&\vdots\vspace*{2pt}
\cr
0&0&0&0&\ldots&\sin(\theta_d)}\right]
.
\end{equation}
Above, the principal angles are set in such a way that $\cos\theta_i$
decreases linearly from $\cos\theta$ to $\alpha\cos\theta$, where
$\theta$ and $\alpha$ are fixed parameters; that is to say,
$\cos\theta_i=(1-a(i-1))\cos\theta$, $a=\frac{1-\alpha}{d-1}$.

\begin{figure}

\includegraphics{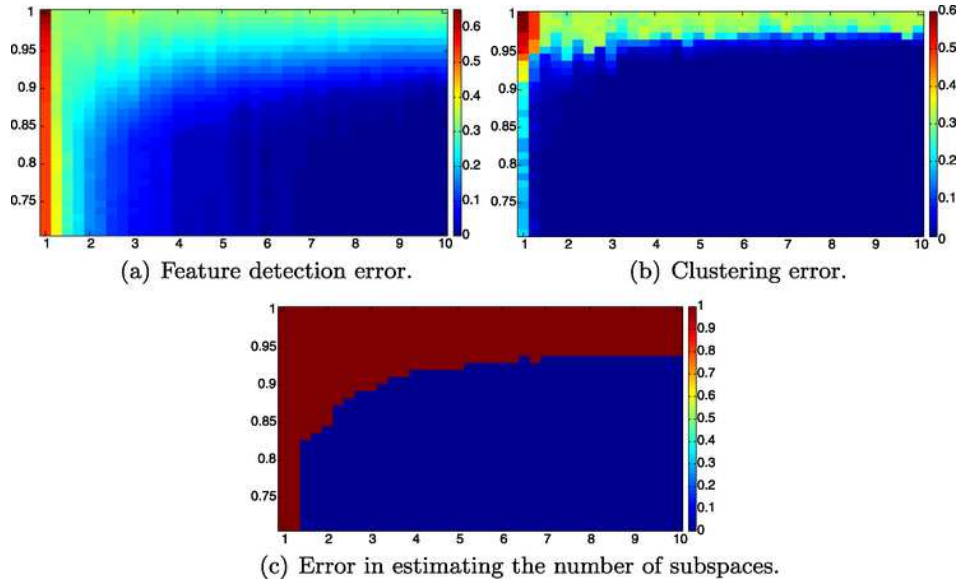}

\caption{Performance of the SSC algorithm for different values of the
affinity and density of points per subspace. In all three figures,
the horizontal axis is the density $\rho$, and the vertical axis is
the normalized maximum affinity $\operatorname{max}_{i \neq j}
\operatorname{aff}(S_i,S_j)/\sqrt{d}$.}\vspace*{-2pt}
\label{figerrcriterion}
\end{figure}

In our experiments we sample $\rho d$ points uniformly at random from
each subspace. We fix $\alpha=\frac{1}{2}$ and vary $\rho\in[2,10]$
and $\theta\in[0,\frac{\pi}{2}]$. Since $\alpha=\frac{1}{2}$, as
$\theta$ increases from $0$ to $\pi/2$, the normalized maximum
affinity $\operatorname{max}_{i \neq j}   \operatorname{aff}(S_i,S_j)/\break\sqrt{d}$
decreases from $1$ to $0.7094$ (recall that a normalized affinity
equal to 1 indicates a perfect overlap, that is, two subspaces are the
same). For each value of $\rho$ and $\theta$, we evaluate the SSC
performance according to the three error criteria above. The results,
shown in Figure~\ref{figerrcriterion}, indicate that SSC is
successful even for large values of the maximum affinity as long as
the density is sufficiently large. Also, the figures display a clear
correlation between the three different error criteria, indicating that
each could be used as a proxy for the other two. An interesting point
is $\rho=3.25$ and $\operatorname{aff}/\sqrt{d}=0.9$; here, the algorithm can
identify the number of subspaces correctly and perform perfect
subspace clustering (clustering error is $0$). This indicates that the
SSC algorithm in its full generality can achieve perfect subspace
clustering even when the subspaces are very close.
%
%f11 #&#

%{
%}
%{
%}
%{
%}
%

%s5.1.4 #&#
\subsubsection{Effect of dimension on subspace detection property and
spectral gap}
\label{dim}
\begin{figure}

\includegraphics{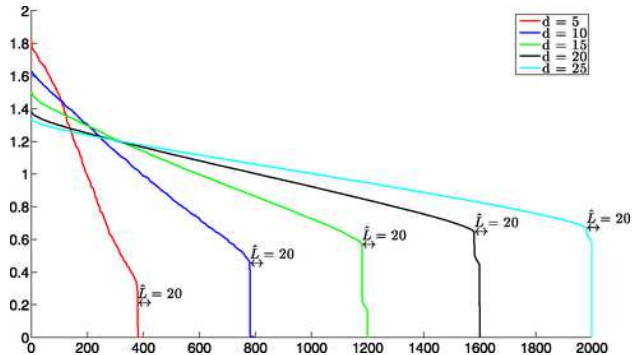}

\caption{Gaps in the eigenvalues of the normalized Laplacian as a
function of subspace dimension.}
\label{figgapsnoiseless}
\end{figure}

In order to illustrate the effect an increase in the dimension of
subspaces has on the spectral gap, we generate $L=20$ subspaces chosen\vadjust{\goodbreak}
uniformly at random from all $d$-dimensional subspaces in
$\mathbb{R}^{50}$. We consider $5$ different values for $d$, namely,
5, 10, 15, 20, 25. In all these cases, the total dimension of the
subspaces $Ld$ is more than the ambient dimension $n=50$. We generate
$4d$ unit-normed points on each subspace uniformly at random. The
corresponding singular values of the normalized Laplacian are
displayed in Figure~\ref{figgapsnoiseless}. As evident from this
figure, the subspace detection property holds, when the dimension of
the subspaces is less than $10$ (this corresponds to the last
eigenvalues being exactly equal to $0$). %Notice that if we exclude the
%log factors \EJC{I don't think we can exclude log factors and
% unspecified constants like this unless we have a precise asymptotic
% theorem. I would remove some of what follows.} Theorem~\ref{th0}
%guarantees that the subspace detection property holds up to $d\le7$,
%verifying the fact that the constants in the prediction of Theorem
Beyond $d=10$, the gap is still
evident, however, the gap decreases as $d$ increases. In all these
cases, the gap was detectable using the sharpest descent heuristic
presented in Algorithm~\ref{algSSC} and, thus, the correct estimates
for the number of subspaces were always found.
%
%f12 #&#

%s5.1.5 #&#
\subsubsection{Effect of noise on spectral gap}
\label{noiseModel}

\begin{figure}

\includegraphics{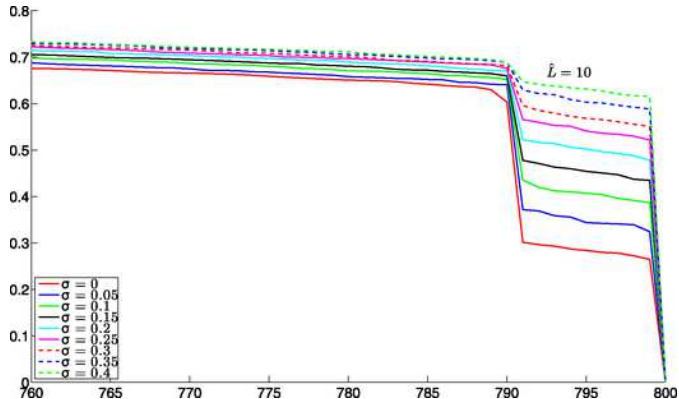}

\caption{Gaps in the eigenvalues of the normalized Laplacian for
different values of the noise level~$\sigma$.}
\label{figgapsnoisy}
\end{figure}

In order to illustrate the effect of noise on the spectral gap, we
sample $L=10$ subspaces chosen uniformly at random from all
$d=20$-dimensional subspaces in $\mathbb{R}^{50}$. The total dimension
of the subspaces ($Ld=200$) is once again more than the ambient
dimension $n=50$. We then sample points on each subspace---$4d$ per
subspace as before---and perturb each unit-norm data point $\vct{x}_i$
by a noisy vector chosen independently and uniformly at random on the
sphere of radius $\sigma$ (noise level) and then normalize to have
unit norm. The noisy samples are
$\tilde{\vct{x}_i}=\frac{\vct{x}_i+\vct{z}_i}{\twonorm{\vct{x}_i+\vct{z}_i}}$,
where $\twonorm{\vct{z}_i} = \sigma$. We consider $9$ different
values for the noise level, namely, 0, 0.05, 0.1, 0.15, 0.2, 0.25,
0.3, 0.35, 0.4. The corresponding singular values of the normalized
Laplacian are shown in Figure~\ref{figgapsnoisy}. As evident from
this figure, we are in a regime where the \sdp\ does not hold even for
noiseless data (this corresponds to the last eigenvalues not being exactly
equal to $0$). For $\sigma$ positive, the gap is still evident but
decreases as a function of $\sigma$. In all these cases, the gap was
detectable using the sharpest descent heuristic presented in Algorithm
\ref{algSSC} and, thus, the number of subspaces was always correctly
inferred.
%
%f13 #&#

\subsubsection{Comparison with other methods}
We now hope to demonstrate that one of the main advantages of
SSC %to classic as well as some recent algorithms
is its ability to identify, in much broader circumstances, the correct
number of subspaces using the eigen-gap heuristic. Before we discuss
the pertaining numerical results, we quickly review a classical method
in subspace clustering \cite{Costeira98}. Start with the rank-$r$ SVD
$\mtx{X}=\mtx{U}\mct{\Sigma}\mtx{V}^T$ of the data matrix and use
$\mtx{W}=\mtx{V}\mtx{V}^T$ as the affinity matrix. (Interestingly, the
nuclear-norm heuristic also results in the same affinity matrix
\cite{Liu10,Favaro11}). It was shown in \cite{Costeira98} that when the
subspaces are independent, the affinity matrix will be block diagonal
and one can thus perform perfect subspace clustering. When the
subspaces are not independent, the affinity matrix may occasionally be
approximately block diagonal as observed empirically in some
particular computer vision applications. In the presence of noise, or
when the independence assumption is violated, various methods have
been proposed to ``clean up'' the affinity matrix and put it into block
diagonal form \cite{Costeira98,Kanatani01,Ichimura99,Wu01,Kanatani02,Kanatani98}. As noted by Vidal in~\cite{Vidal11}, most of
these algorithms need some knowledge of the true data rank and/or
dimension of the subspaces. Furthermore, none of these algorithms have
been proven to work when the independence criterion is violated---in
contrast with the analysis presented in this paper.

%From a more practical viewpoint, one major drawback of all these
%algorithms (including SSC when the Subspace Detection Property is
%violated) is that the number of subspaces has to somehow be estimated
%using the data. As mentioned a good method to estimate the number of
%subspaces is to use the eigen-gap heuristic.

\begin{figure}

\includegraphics{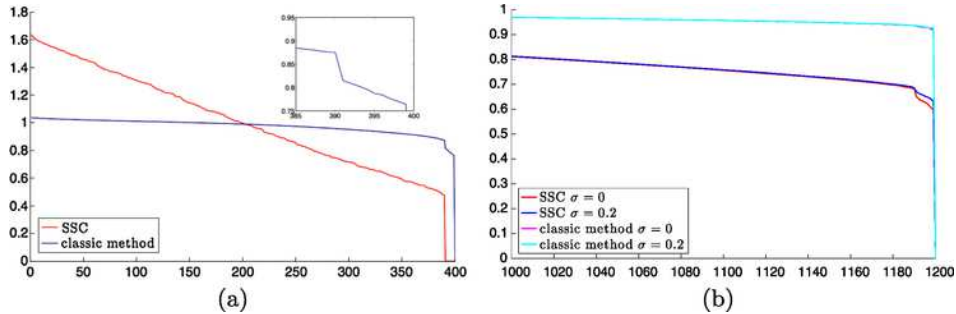}

\caption{Gaps in the eigenvalues of the normalized Laplacian for the
affinity graphs. \textup{(a)} Noiseless setup with $d = 10$
(the zoom is to see the gap for the classical method
more clearly). \textup{(b)} Noiseless and noisy setups with $d = 30$. }
\label{figsimpleest}
\end{figure}

We believe that a major advantage of SSC vis a vis more recent
approaches \cite{Liu10,Favaro11} is that the eigen-gap heuristic is
applicable under broader circumstances. To demonstrate this, we sample
$L=10$ subspaces chosen uniformly at random from all $10$-dimensional
subspaces in $\mathbb{R}^{50}$. The total dimension $Ld = 100$ is once
more larger than the ambient dimension $n = 50$. The eigenvalues of
the normalized Laplacian of the affinity matrix for both SSC and the
classical method ($\mtx{W}=\mtx{V}\mtx{V}^T$) are shown in Figure
\ref{figsimpleest}(a). Observe that the gap exists in both
plots. However, SSC demonstrates a wider gap and, therefore, the
estimation of the number of subspaces is more robust to noise. To
illustrate this point further, consider Figure~\ref{figsimpleest}(b)
in which points are sampled according to the same scheme but with
$d = 30$, and with noise possibly added just as in Section
\ref{noiseModel}. Both in the noisy and noiseless cases, the classical
method does not produce a detectable gap, while the gap is detectable
using the simple methodology presented in Algorithm~\ref{algSSC}.
%
%f14 #&#

%f15 #&#
\begin{figure}

\includegraphics{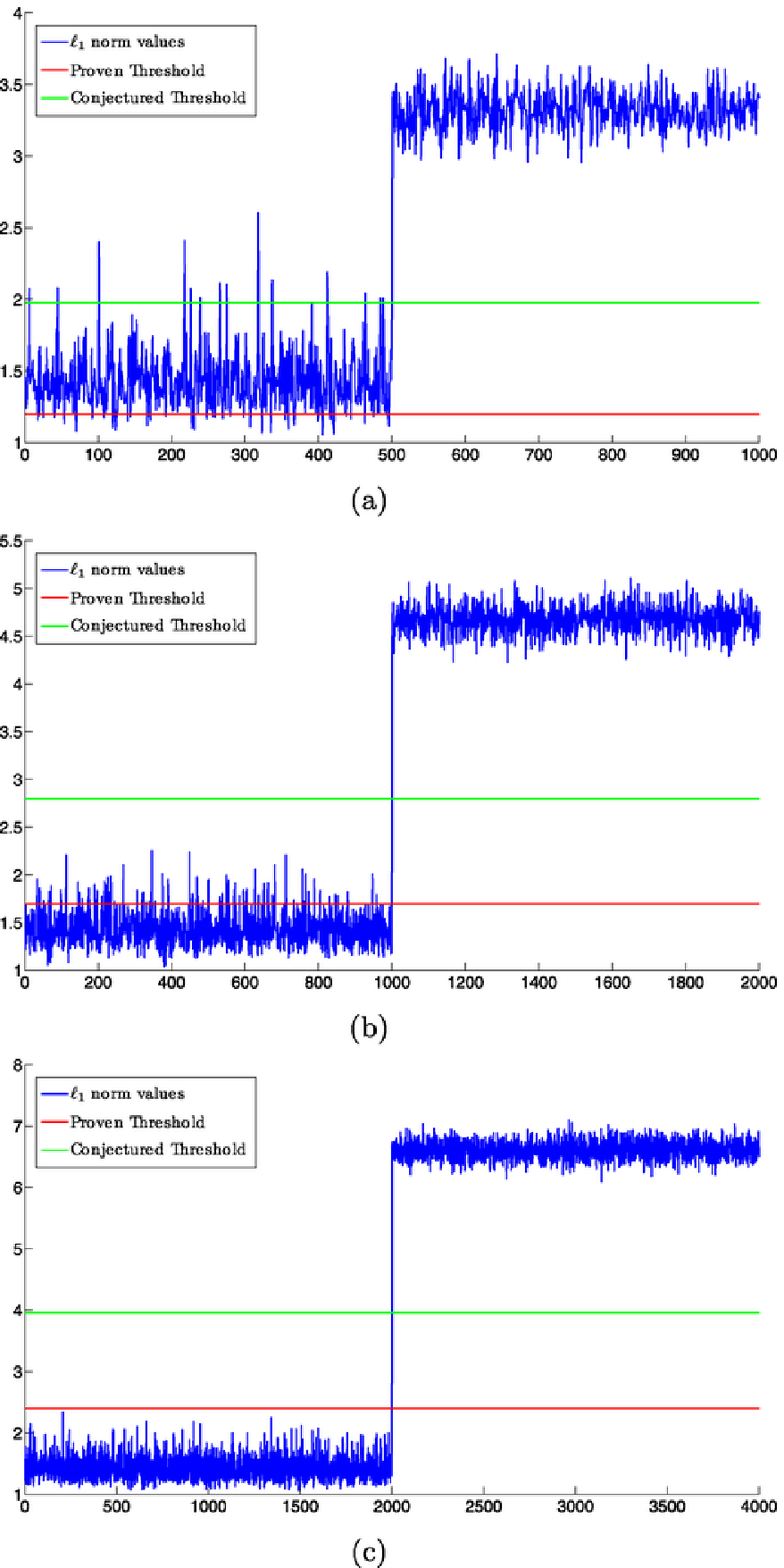}

\caption{Gap in the optimal values with $L=2{n}/{d}$ subspaces. \textup{(a)}
$d=5$, $n=50$, $L = 20$. \textup{(b)} $d = 5$, $n = 100$, $L = 40$. \textup{(c)}
$d=5$, $n=200$, $L = 80$.}
\label{figoutlier1}
\end{figure}

%To summarize, when outliers are not present, SSC has performance
%comparable with the state of the art \cite{Favaro11} when the number
%of subspaces are known in advance. In the case of corrupted
%trajectories as demonstrated by \cite{Favaro11}, SSC has the best
%performance among existing algorithms. Also as shown in this paper SSC
%is rigorously provable under broader circumstances. From a practical
%viewpoint when the trajectories are not corrupted the major drawback
%of SSC in comparison with \cite{Favaro11} is that it is
%computationally less efficient. However, it is computationally more
%efficient when corrupted trajectories are present.

%s5.2 #&#
\subsection{Segmentation with outliers}
We now turn to outlier detection. For this purpose, we consider three
different setups in which
\begin{itemize}
\item$d=5$, $n = 50$,
\item$d = 5$, $n = 100$,
\item$d = 5$, $n = 200$.
\end{itemize}
In each case, we sample $L=2{n}/{d}$ subspaces chosen uniformly at
random so that the total dimension $Ld=2n$. For each subspace, we
generate $5d$ points uniformly at random so that the total number of
data points is $N_d= 10n$. We add $N_0= N_d$ outliers chosen uniformly
at random on the sphere. Hence, the number of outliers is equal to the
number of data points. The optimal values of the optimization problems
(\ref{eqel2}) are plotted in Figure~\ref{figoutlier1}. The first
$N_d$ values correspond to the data points and the next $N_0$ values
to the outliers. As can be seen in all the plots, a gap appears in the
values of the $\ell_1$ norm of the optimal solutions. That is, the
optimal value for data points is much smaller than the corresponding
optimal value for outlier points. We have argued that the critical
parameter for outlier detection is the ratio $d/n$. The smaller, the
better. As can be seen in Figure~\ref{figoutlier1}(a), the ratio
$d/n = 1/10$ is already small enough for the conjectured threshold of
Algorithm~\ref{algoutlier} to work and detect all outlier points
correctly. However, it wrongfully considers a few data points as
outliers. In Figure~\ref{figoutlier1}(b), $d/n = 1/20$ and the
conjectured threshold already works perfectly, but the proven threshold
is still not able to do outlier detection well. In
Figure~\ref{figoutlier1}(c), $d/n = 1/40$, both the conjectured and
proven thresholds can perform perfect outlier detection. (In practice,
it is of course not necessary to use the threshold as a criterion for
outlier detection; one can instead use a gap in the optimal
values.) %The surprising conclusion of these numerical examples is that
%even though we oversample each subspace by only a factor of $5$ ($5d$
%points on each subspace), and we have the same number of data points
%as the number of outliers, we can still detect all outliers correctly
%by exploiting this gap.
It is also worth mentioning that if $d$ is larger, the optimal value
is more concentrated for the data points and, therefore, both the
proven and conjectured threshold would work for smaller ratios of
${d}/{n}$ (this is different from the small values of $d$
above). %\EJC{Check that the last sentence makes sense.}
%

%s6 #&#
\section{Background on Geometric Functional Analysis}

Our proofs rely heavily on techniques from Geometric Functional
Analysis and we now introduce some basic concepts and results from this
field. Most of our exposition is adapted from~\cite{Vershynin11}.

%de6.1 #&#
\begin{definition}
The maximal and average values of $\| \cdot\|_\mathcal{K}$ on the sphere
$\mathcal{S}^{n-1}$ are defined by
\[
b(\mathcal{K})= \mathop{\sup}_{\vct{x}\in S^{n-1}} \|\vct{x}\|_\mathcal{K}
\quad\mbox{and}\quad M(\mathcal{K})=\int_{S^{n-1}}\|\vct{x}
\|_\mathcal{K} \,d\sigma (\vct{x}).
\]
Above, $\sigma$ is the uniform probability measure on the sphere.
\end{definition}

%de6.2 #&#
\begin{definition}
The \textit{mean width} $M^*(\mathcal{K})$ of a symmetric convex body
$\mathcal{K}$ in $\mathbb{R}^n$ is the expected value of the dual norm
over the unit sphere,
\[
M^*(\mathcal{K})=M \bigl(\mathcal{K}^o \bigr)=\int
_{\mathcal{S}^{n-1}}\|\vct{y}\|_{\mathcal{K}^o} \,d\sigma(\vct{y})=\int
_{\mathcal{S}^{n-1}} \mathop{\max}_{\vct{z} \in\mathcal{K}}\langle\vct{y},\vct{z}
\rangle \,d\sigma(\vct{y}).
\]
\end{definition}
With this in place, we now record some useful results.

%le6.3 #&#
\begin{lemma}\label{M} We always have
$M(\mathcal{K})M(\mathcal{K}^o)\ge1$.
\end{lemma}
\begin{pf}
Observe that since $\|\cdot\|_{\mathcal{K}^o}$ is the dual norm of
$\|\cdot\|_\mathcal{K}$, $\|\vct{x}\|^2 = \|\vct{x}\|_{\mathcal{K}}
\|\vct{x}\|_{\mathcal{K}^o}$ and, thus,
\[
1 = \biggl(\int_{\mathcal{S}^{n-1}} \sqrt{\|\vct{x}\|_{\mathcal{K}} \|
\vct{x}\|_{\mathcal{K}^o}} \,d\sigma \biggr)^2 \le\int
_{\mathcal{S}^{n-1}} \|\vct{x}\|_{\mathcal{K}} \,d\sigma \int
_{\mathcal{S}^{n-1}} \|\vct{x}\|_{\mathcal{K}^o} \,d\sigma,
\]
where the inequality follows from Cauchy-Schwarz.
\end{pf}
The following theorem deals with concentration properties of
norms. According to \cite{Klartag07}, these appear in the first pages
of \cite{Milman86}.
%
%th6.4 #&#
\begin{theorem}[(Concentration of measure)]
\label{concentration}
For each $t > 0$, we have
\[
\sigma \bigl\{\vct{x}\in S^{n-1} \dvtx \bigl| \|\vct{x}\|_\mathcal
{K}-M(\mathcal{K}) \bigr| > t M(\mathcal{K}) \bigr\} < \exp \biggl(-ct^2n
\biggl[\frac{M(\mathcal{K})}{b(\mathcal{K})} \biggr]^2 \biggr),
\]
where $c>0$ is a universal
constant.% and $k(K)=n{\big(\frac{M(K)}{b(K)}\big)}^2$ is the Dvoretzky
%dimension.
\end{theorem}
The following lemma is a simple modification of a well-known result in
Geometric Functional Analysis.
%
%le6.5 #&#
\begin{lemma}[(Many faces of convex symmetric polytopes)]\label{faces}
Let $\mathcal{P}$ be a symmetric polytope with $f$ faces. Then
\[
n { \biggl(\frac{M(\mathcal{P})}{b(\mathcal{P})} \biggr)}^2 \le c \log(f)
\]
for some positive numerical constant $c > 0$.
\end{lemma}
%
%de6.6 #&#
\begin{definition}[(Geometric banach-mazur distance)]\label{BM}
Let $\mathcal{K}$ and $\mathcal{L}$ be symmetric convex bodies in
$\mathbb{R}^n$. The
Banach-Mazur distance between $\mathcal{K}$ and $\mathcal{L}$, denoted
by $d(\mathcal{K},\mathcal{L})$, is the least positive
value $a b \in\R$ for which there is a linear image $T(\mathcal{K})$
of $\mathcal{K}$
obeying
\[
b^{-1} \mathcal{L} \subseteq T(\mathcal{K}) \subseteq a \mathcal{L}.
\]
\end{definition}
%
%th6.7 #&#
\begin{theorem}[(John's theorem)]
Let $\mathcal{K}$ be a symmetric convex body in $\mathbb{R}^n$ and
$B_2^n$ be
the unit ball of $\mathbb{R}^n$. Then $d(\mathcal{K},B_2^n)\le\sqrt{n}$.
\end{theorem}

Our proofs make use of two theorems concerning volume ratios. The
first is this.
%
%le6.8 #&#
\begin{lemma}[(Urysohn's inequality)]\label{Urysohn} Let $\mathcal{K}
\subset
\mathbb{R}^n$ be a compact set. Then
\[
{ \biggl(\frac{\operatorname{{vol}}(\mathcal{K})}{\operatorname{vol}(B_2^n)} \biggr)}^{{1}/{n}} \le M^*(\mathcal{K}).
\]
\end{lemma}

%le6.9 #&#
\begin{lemma}[(\cite{Ball90}, Theorem 2)]
\label{Ball}
Let $\mathcal{K}^o= \{\vct{z}\in\mathbb{R}^n \dvtx  | \langle\vct{a}_i ,
\vct{z} \rangle|\le1 \dvtx  i = 1, \ldots, N\}$ with
$\twonorm{\vct{a}_i}=1$. The volume of $\mathcal{K}^o$ admits the lower estimate
\[
{\operatorname{{vol}} \bigl(\mathcal{K}^o
\bigr)}^{1/n} \ge %
\cases{\displaystyle \frac{2\sqrt{2}}{\sqrt{p}r}, &\quad $p\ge2,$
\vspace*{2pt}
\cr
\displaystyle\frac{1}{r}, &\quad $\mbox{if } 1\le p \le2.$} %
\]
Here, $n\le N$, $1 \le p <\infty$ and $r={ (\frac{1}{n}\sum_{i =
1}^{N}\twonorm{\vct{a}_i}^p )}^{{1}/{p}}$.
\end{lemma}

%s7 #&#
\section{Proofs}
\label{proofs}

%With our notation, the total number of points obeys: $N = N_0 + N_1 +
%$\mtx{X} \in\mathbb{R}^{n\times N}$. We can represent this matrix
%as $\mtx{X}=[\mtx{X^{(0)}}, \mtx{X^{(1)}}, \ldots,
%$N_\ell$ columns of $\mtx{X^{(\ell)}}\in\mathbb{R}^{n\times
%N_\ell}$ are those vectors in $\mathcal{X}^{(\ell)}$. It is no loss
%of generality to suppose that $\mct{\Gamma}=\mtx{I}$ (of course
%we do not know the ordering in practice). Finally, $\vct{x_i}$ is the
%$i$th data point --- the $i$th column of $\mtx{X}$ --- and we also set
%$\mtx{X}_{-i} =
%[\vct{x}_1,\ldots,\vct{x}_{i-1},\vct{x}_{i}, \ldots,
%column.

To avoid repetition, we define the primal optimization problem $P(\vct
{y},\mtx{A})$ as
%
%e7.1 #&#
\[
\mathop{\min}_{\vct{x}} \|\vct{x}\|_{\ell_1}\qquad \mbox{subject to }
\mtx{A}\vct{x} = \vct{y},
\]
and its dual $D(\vct{y},\mtx{A})$ as
%
%e7.2 #&#
\[
\mathop{\max}_{\mct{\nu}} \langle\vct{y},\mct{\nu}\rangle \qquad\mbox{subject to }
\bigl\|\mtx{A}^T\mct{\nu}\bigr\|_{\ell_\infty} \le1.
\]
We denote the optimal solutions by $\operatorname{optsolP}(\vct{y},\mtx{A})$
and $\operatorname{optsolD}(\vct{y},\mtx{A})$. Since the primal is a linear
program, strong duality holds, and both the primal and dual have the
same optimal value which we denote by $\operatorname{optval}(\vct{y},\mtx{A})$
(the optimal value is set to infinity when the primal problem is
infeasible). Also notice that as discussed in Section~\ref{Geompers},
this optimal value is equal to $\|\vct{y}\|_{\mathcal{K}}$, where
$\mathcal{K}(\mtx{A})=\operatorname{conv}(\pm\vct{a}_1,\ldots,\pm\vct{a}_N)$
and $\mathcal{K}^o(\mtx{A})=\{\vct{z}\dvtx  \infnorm{\mtx{A}^T\vct{z}}\le1\}$.

%s7.1 #&#
\subsection{\texorpdfstring{Proof of Theorem \protect\ref{th1}}{Proof of Theorem 2.5}}\label{prfth1}
We first prove that the geometric condition (\ref{GeomCond}) implies
the subspace detection property. We begin by establishing a simple
variant of a now classical lemma (e.g., see \cite{Candes06}). Below, we
use the notation $\mtx{A}_S$ to denote the submatrix of $\mtx{A}$ with
the same rows as $\mtx{A}$ and columns with indices in $S \subset\{1,
\ldots, N\}$.
%
%le7.1 #&#
\begin{lemma}
\label{dualcertificate}
Consider a vector $\vct{y}\in\R^n$ and a matrix $\mtx{A}\in\R^{n\times
N}$. If there exists~$\vct{c}$ obeying $\vct{y}=\mtx{A}\vct{c}$ with
support $S\subseteq T$, and a dual certificate vector $\mct{\nu}$
satisfying
\[
\mtx{A}_{S}^T\mct{\nu}=\sgn{\vct{c}_{S}}, \qquad \bigl\|
\mtx{A}_{T\cap S^c}^T\mct{\nu}\bigr\|_{\ell_\infty} \le1, \qquad \bigl\|
\mtx{A}_{T^c}^T\mct{\nu}\bigr\|_{\ell_\infty} < 1,
\]
then all optimal solutions $\vct{z}^*$ to $P(\vct{y},\mtx{A})$ obey
$\vct{z}_{T^c}^*=\vct{0}$.
\end{lemma}
\begin{pf}
Observe that for any optimal solution $\vct{z}^*$ of $P(\vct{y},\mtx
{A})$, we have
\begin{eqnarray*}
\bigl\|\vct{z}^*\bigr\|_{\ell_1} &=& \bigl\|\vct{z}_{S}^*\bigr\|_{\ell_1} +
\bigl\|\vct{z}_{T\cap
S^c }^*\bigr\|_{\ell_1} + \bigl\|\vct{z}_{T^c}^*
\bigr\|_{\ell_1}
\\
&\ge& \|\vct{c}_{S}\|_{\ell_1} + \bigl\langle\sgn{
\vct{c}_S}, \vct{z}_{S}^*-\vct {c}_{S} \bigr
\rangle+ \bigl\|\vct{z}_{T\cap S^c }^*\bigr\|_{\ell_1} + \bigl\|\vct
{z}_{T^c}^*\bigr\|_{\ell_1}
\\
&=& \|\vct{c}_{S}\|_{\ell_1} + \bigl\langle\mct{\nu},
\mtx{A}_S \bigl(\vct{z}_{S}^*-\vct {c}_{S}
\bigr) \bigr\rangle+ \bigl\|\vct{z}_{T\cap S^c }^*\bigr\|_{\ell_1} + \bigl\|\vct
{z}_{T^c}^*\bigr\|_{\ell_1}
\\
&=& \|\vct{c}_{S}\|_{\ell_1} + \bigl\|\vct{z}_{T\cap S^c }^*
\bigr\|_{\ell_1} - \bigl\langle\mct {\nu},\mtx{A}_{T\cap S^c}
\vct{z}_{T\cap S^c}^{*} \bigr\rangle+ \bigl\|\vct {z}_{T^c}^*
\bigr\|_{\ell_1} - \bigl\langle\mct{\nu}, \mtx{A}_{T^c}
\vct{z}_{T^c}^* \bigr\rangle .
\end{eqnarray*}
Now note that
\[
\bigl\langle\mct{\nu},\mtx{A}_{T\cap S^c}\vct{z}_{T\cap S^c}^* \bigr
\rangle= \bigl\langle \mtx{A}_{T\cap S^c}^T\mct{\nu},
\vct{z}_{T\cap S^c}^* \bigr\rangle\le \bigl\|\mtx{A}_{T\cap S^c}^T
\mct{\nu}\bigr\|_{\ell_\infty} \bigl\|\vct{z}_{T\cap S^c}^*\bigr\|_{\ell_1} \le \bigl\|
\vct{z}_{T\cap S^c}^*\bigr\|_{\ell_1} .
\]
In a similar manner, we have $\langle\mct{\nu},\mtx{A}_{T^c}\vct
{z}_{T^c}^*\rangle\le\infnorm{\mtx{A}_{T^c}^T\mct{\nu}}\onenorm{\vct
{z}_{T^c}^*}$. Hence, using these two identities, we get
\[
\bigl\|\vct{z}^*\bigr\|_{\ell_1} \ge \|\vct{c}\|_{\ell_1} + \bigl(1- \bigl\|
\mtx{A}_{T^c}^T \mct {\nu}\bigr\|_{\ell_\infty} \bigr) \bigl\|
\vct{z}_{T^c}^*\bigr\|_{\ell_1} .
\]
Since $\vct{z}^*$ is an optimal solution, $\onenorm{\vct{z}^*}\le
\onenorm{\vct{c}}$, and plugging this into the last identity gives
\[
\bigl(1- \bigl\|\mtx{A}_{T^c}^T\mct{\nu}\bigr\|_{\ell_\infty}
\bigr) \bigl\|\vct{z}_{T^c}^{*}\bigr\|_{\ell_1} \le 0.
\]
Now since $\infnorm{\mtx{A}_{T^c}^T\mct{\nu}}<1$, it follows that
$\onenorm{\vct{z}_{T^c}^*}=0$.
\end{pf}
Consider $\vct{x}_{i}^{(\ell)}=\mtx{U}^{(\ell)}\vct{a}_{i}^{(\ell)}$,
where $\mtx{U}^{(\ell)}\in\R^{n\times d_\ell}$ is an orthogonal basis
for $S_\ell$ and define
%
%e7.3 #&#
\[
\vct{c}_{i}^{(\ell)}=\operatorname{optsolP} \bigl(
\vct{a}_{i}^{(\ell)},\mtx {A}_{-i}^{(\ell)}
\bigr).
\]
Letting $S$ be the support of $\vct{c}_i^{(\ell)}$, define
$\mct{\lambda}_{i}^{(\ell)}$ as an optimal solution to
\begin{eqnarray}
\mct{\lambda}_{i}^{(\ell)}=\mathop{\arg\min}_{\bar{\mct{\lambda}}_{i}^{(\ell)}\in
\R^{d_\ell}}
\bigl\|\bar{\mct{\lambda}}_{i}^{(\ell)}\bigr\|_{\ell_2}\nonumber\\
\eqntext{\mbox{subject to } \bigl\{{ \bigl(\mtx{A}_{-i}^{(\ell)}
\bigr)}_{S}^T \bar{\mct{\lambda}}_{i}^{(\ell)}=
\operatorname{sgn}\bigl(\vct{c}_{i}^{(\ell)}\bigr), \bigl\|{ \bigl(\mtx{A}_{-i}^{(\ell)}
\bigr)}_{S^c}^T \bar{\mct{\lambda }}_{i}^{(\ell)}
\bigr\|_{\ell_\infty} \le1 \bigr\}.}
\end{eqnarray}
Because $\vct{c}_i^{(\ell)}$ is optimal for the primal problem, the
dual problem is feasible by strong duality and the set above is
nonempty. Also, $\mct{\lambda}_{i}^{(\ell)}$ is a dual point in the
sense of Definition~\ref{dualpointdef},
that is, $\mct{\lambda}_i^{(\ell)}=\mct{\lambda}(\vct{a}_i^{(\ell)},\mtx
{A}_{-i}^{(\ell)})$. Introduce
\[
\mct{\nu}_i^{(\ell)}=\mtx{U}^{(\ell)}\vct{
\lambda}_i^{(\ell)},
\]
so that the direction of $\mct{\nu}_i^{(\ell)}$ is the $i$th dual
direction, that is, $\mct{\nu}_i^{(\ell)}=\twonorm{\mct{\lambda
}_i^{(\ell)}}\vct{v}_{i}^{(\ell)}$ (see Definition~\ref{dualdirectiondef}).

Put $T$ to index those columns of $\mtx{X}_{-i}$ in the same subspace
as $\vct{x}_i^{(\ell)}$ (subspace $S_\ell$). Using this definition,
the \sdp\ holds if we can prove the existence of vectors $\vct{c}$
(obeying $\vct{c}_{T^c}=\vct{0}$) and $\mct{\nu}$ as in Lemma~\ref{dualcertificate} for problems
$P (\vct{x}_i^{(\ell)},\mtx{X}_{-i} )$ of the form
%
%e7.4 #&#
\begin{equation}
\label{mainopt} \min_{\vct{z} \in\R^{N-1}} \|\vct{z}\|_{\ell_1} \qquad\mbox {subject
to } \mtx{X}_{-i}\vct{z} = \vct{x}_{i}^{(\ell)}.
\end{equation}
%
% \vct{x}_{i}^{(\ell)} = \begin{bmatrix}\mtx{X}^{(1)},\ldots,
%satisfies $\vct{c}^{*(k)}=\vct{0}$ for all $k=1,\ldots,\ell-1,\ell+1,
%Based on Lemma~\ref{dualcertificate}, it is sufficient to prove that
%for to finding a dual certificate $\mct{\nu}_i^{(\ell)}\in\R^n$, such
%that
We set to prove that the vectors
$\vct{c}=
\pmatrix{\vct{0},\ldots,\vct{0},\vct{c}_i^{(\ell)},\vct{0},\ldots
,\vct{0}
}
$,
which obeys $\vct{c}_{T^c}=\vct{0}$ and is feasible for
\eqref{mainopt}, and $\mct{\nu}_i^{(\ell)}$ are indeed as in Lemma
\ref{dualcertificate}. To do this, we have to check that the
following conditions are satisfied:
%
%e7.5 #&#
\begin{eqnarray}
\label{eqcond} { \bigl({\mtx{X}_{-i}^{(\ell)}}
\bigr)}_{S}^T\mct{\nu}_i^{(\ell)}&=&\operatorname{sgn}\bigl(
\vct {c}_{i}^{(\ell)}\bigr),
\\
\label{ineqcond1} \bigl\|{ \bigl({\mtx{X}_{-i}^{(\ell)}}
\bigr)}_{S^c}^T\mct{\nu}_i^{(\ell
)}
\bigr\|_{\ell_\infty} &\le&1,
\end{eqnarray}
and for all $\vct{x}\in\mathcal{X}\setminus\mathcal{X}_\ell$
%
%e7.7 #&#
\begin{equation}
\label{ineqcond2} \bigl| \bigl\langle\vct{x},\mct{\nu}_i^{(\ell)}
\bigr\rangle\bigr|<1.
\end{equation}
Conditions (\ref{eqcond}) and (\ref{ineqcond1}) are satisfied by
definition, since
%
%e7.8 #&#
\[
{ \bigl({\mtx{X}_{-i}^{(\ell)}} \bigr)}_{S}^T
\mct{\nu}_i^{(\ell)}={ \bigl({\mtx{A}_{-i}^{(\ell)}}
\bigr)}_{S}^T{\mtx{U}^{(\ell)}}^T
\mtx{U}^{(\ell
)}\mct{\lambda}_{i}^{(\ell)}={ \bigl({
\mtx{A}_{-i}^{(\ell)}} \bigr)}_{S}^T\vct{
\lambda}_{i}^{(\ell)}=
\operatorname{sgn}\bigl(\vct{c}_{i}^{(\ell)}\bigr),
\]
and
%
%e7.9 #&#
\[
\bigl\|{ \bigl({\mtx{X}_{-i}^{(\ell)}} \bigr)}_{S^c}^T
\mct{\nu}_i^{(\ell
)}\bigr\|_{\ell_\infty} = \bigl\|{ \bigl({
\mtx{A}_{-i}^{(\ell)}} \bigr)}_{S^c}^T{
\mtx{U}^{(\ell
)}}^T \mtx{U}^{(\ell)}\mct{
\lambda}_{i}^{(\ell)}\bigr\|_{\ell_\infty} = \bigl\|{ \bigl({\mtx
{A}_{-i}^{(\ell)}} \bigr)}_{S^c}^T\mct{\lambda}_{i}^{(\ell)}\bigr\|_{\ell_\infty}\le 1.
\]
Therefore, in order to prove that the \sdp\ holds, it remains to check
that for all $\vct{x}\in\mathcal{X}\setminus\mathcal{X}_\ell$ we have
%
%e7.10 #&#
\[
\bigl|\bigl\langle\vct{x},\mct{\nu}_i^{(\ell)} \bigr\rangle\bigr|=
\bigl| \bigl\langle\vct {x},\vct{v}_i^{(\ell)} \bigr\rangle\bigr| \bigl\|
\mct{\lambda}_i^{(\ell
)}\bigr\|_{\ell_2} <1.
\]
By definition of $\mct{\lambda}_i^{(\ell)}$, $\infnorm{{\mtx
{A}_{-i}^{(\ell)}}^T\mct{\lambda}_i^{(\ell)}}\le1$ and, therefore, $\vct
{\lambda}_i^{(\ell)}\in{ (\mathcal{P}_{-i}^{\ell} )}^o$, where
%
%e7.11 #&#
\[
{ \bigl(\mathcal{P}_{-i}^{\ell} \bigr)}^o= \bigl\{
\vct{z}\dvtx \bigl\|{\mtx {A}_{-i}^{(\ell)}}^T \vct{z}
\bigr\|_{\ell_\infty} \le1 \bigr\}.
\]

%de7.2 #&#
\begin{definition}[(Circumradius)]
The circumradius of a convex body $\mathcal{P}$, denoted by $R(\mathcal
{P})$, is defined as the radius of the smallest ball containing
$\mathcal{P}$.
\end{definition}
Using this definition and the fact that $\mct{\lambda}_i^{\ell}\in{
(\mathcal{P}_{-i}^{\ell} )}^o$, we have
%
%e7.12 #&#
\[
\bigl\|\mct{\lambda}_i^{(\ell)}\bigr\|_{\ell_2} \le R \bigl({
\mathcal{P}_{-i}^{\ell
}}^o \bigr)=
\frac{1}{r({\mathcal{P}_{-i}^{\ell}})},
\]
where the equality is a consequence of the lemma below.
%
%le7.3 #&#
\begin{lemma}[(\cite{Brandenberg04}, page 448)]
\label{invnormlemma}
For a symmetric convex body $\mathcal{P}$, that is, $\mathcal
{P}=-\mathcal{P}$, the following relationship between the inradius of
$\mathcal{P}$ and circumradius of its polar $\mathcal{P}^o$ holds:
%
%e7.13 #&#
\[
r(\mathcal{P})R \bigl(\mathcal{P}^o \bigr)=1.
\]
\end{lemma}
In summary, it suffices to verify that for all pairs $(\ell,i)$ (a
pair corresponds to a point $\vct{x}_i^{(\ell)} \in\mathcal{X}_\ell$)
and all $\vct{x}\in\mathcal{X}\setminus\mathcal{X}_\ell$, we have
\[
\bigl|\bigl\langle\vct{x},\vct{v}_i^{(\ell)} \bigr\rangle\bigr|<r
\bigl(\mathcal{P}_{-i}^\ell \bigr).
\]
Now notice that the latter is precisely the sufficient condition given
in the statement of Theorem~\ref{th1}, thereby concluding the proof.

%s7.2 #&#
\subsection{\texorpdfstring{Proof of Theorem \protect\ref{th2}}{Proof of Theorem 2.8}}
We prove this in two steps.
\begin{longlist}
\item[\textit{Step 1}:] We develop a lower bound about the inradii, namely,
%
%e7.14 #&#
\begin{equation}
\label{step1boundonrad}\qquad \mathbb{P} \biggl\{\frac{c(\rho_\ell)\sqrt{\log\rho_\ell}}{\sqrt{2d_\ell
}}\le r \bigl(
\mathcal{P}^\ell_{-i} \bigr) \mbox{ for all pairs } (\ell,i)
\biggr\} \ge1-\sum_{\ell=1}^L N_\ell
e^{-\sqrt{\rho_\ell}d_\ell}.
\end{equation}
\item[\textit{Step 2}:] Notice that $\mu(\mathcal{X}_\ell)=\max_{k
:   k\neq
\ell}\infnorm{{\mtx{X}^{(k)}}^T\mtx{V}^{(\ell)}}$. %\footnote{We
% remind the reader that we use $\infnorm{\cdot}$ for a matrix to
% mean maximum absolute value of its entries.}
Therefore, we develop
an upper bound about the subspace incoherence, namely,
%
%e7.15 #&#
%e7.16 #&#
\begin{eqnarray}
\label{step2boundinco} \mathbb{P} \biggl\{ \bigl\|{\mtx{X}^{(k)}}^T
\mtx{V}^{(\ell)}\bigr\|_{\ell_\infty} &\le&4 \bigl(\log \bigl[N_\ell(N_k+1)
\bigr] + \log L+t \bigr)\frac{\operatorname{aff}(S_k,S_\ell
)}{\sqrt{d_k}\sqrt{d_\ell}}\nonumber\\
&&\hspace*{62pt}{} \mbox{for all pairs } (\ell, k)
\mbox{ with } \ell\neq k \biggr\}
\\
&\ge& 1-\frac{1}{L^2}\sum_{k\neq\ell}
\frac{4}{(N_k+1)N_\ell}e^{-2t}.\nonumber
\end{eqnarray}
\end{longlist}
Notice that if the condition (\ref{affcond}) in Theorem~\ref{th2}
holds, that is,
\[
\mathop{\max}_{k\neq\ell} 4\sqrt{2} \bigl(\log \bigl[N_\ell(N_k+1)
\bigr] +\log L+t \bigr)\frac{\operatorname{aff}(S_k,S_\ell)}{\sqrt{d_k}}<c(\rho_\ell )\sqrt{\log
\rho_\ell},
\]
then steps~$1$ and~$2$ imply that the deterministic condition in
Theorem~\ref{th1} holds with high probability. In turn, this gives the
subspace detection property.

%s7.2.1 #&#
\subsubsection{Proof of step 1}
Here, we simply make use of a lemma stating that the inradius of a
polytope with vertices chosen uniformly at random from the unit sphere
is lower bounded with high probability.
%
%le7.4 #&#
\begin{lemma}[(\cite{Alonso08})]
\label{inradius}
Assume $\{P_i\}_{i=1}^N$ are independent random vectors on
$\mathbb{S}^{d-1}$, and set $\mathcal{K}=\operatorname{{conv}}(\pm P_1,
\ldots, \pm P_N)$. For every $\delta> 0$, there exists a constant
$C(\delta)$ such that if $(1+\delta)d<N<de^{{d}/{2}}$, then
%
%e7.17 #&#
\[
\mathbb{P} \biggl\{ r(\mathcal{K})<\min \bigl\{C(\delta),1/\sqrt{8} \bigr\}
\sqrt{ \frac{\log({N}/{d})}{d}} \biggr\} \le e^{-d}.
\]
Furthermore, there exists a numerical constant $\delta_0$ such that for
all $N>d(1+\delta_0)$ we have
%
%e7.18 #&#
\[
\mathbb{P} \biggl\{ r(\mathcal{K})< \frac{1}{\sqrt{8}}\sqrt{\frac{\log({N}/{d})}{d}}
\biggr\} \le e^{-d}.
\]
\end{lemma}
One can increase the probability with which this lemma holds by
introducing a parameter $0 < \beta\le1$ in the lower bound \cite
{Gluskin88}. A
modification of the arguments yields (note the smaller bound on the
probability of failure)\looseness=1
%
%e7.19 #&#
\[
\mathbb{P} \biggl\{r(\mathcal{K})<\min \bigl\{C(\delta),{1}/{\sqrt{8}} \bigr\}
\sqrt {\beta\frac{\log({N}/{d})}{d}} \biggr\}\le e^{-d^\beta N^{1-\beta
}}.
\]\looseness=0
This is where the definition of the constant
$c(\rho)${\footnote{Recall that $c(\rho)$ is defined as a constant
obeying the following two properties: (i) for all $\rho> 1$,
$c(\rho)>0$; (ii) there is a numerical value $\rho_0$, such that
for all $\rho\ge\rho_0$, one can take
$c(\rho)=\frac{1}{\sqrt{8}}$.} comes in. We
set $c(\rho)=\min\{C(\rho-1),1/\sqrt{8}\}$ and $\rho_0=\delta_0+1$
where $\delta_0$ is as in the above Lemma and use $\beta=\frac{1}{2}$.
Now since
$\mathcal{P}_{-i}^\ell$ consists of $2(N_\ell-1)$ vertices on
$\mathbb{S}^{d_\ell-1}$ taken from the intersection of the unit sphere
with the subspace $S_\ell$ of dimension $d_\ell$, applying Lemma
\ref{inradius} and using the union bound establishes
\eqref{step1boundonrad}.
%Applying the union bound again, implies
%s7.2.2 #&#
\subsubsection{Proof of step 2}
By definition,
%
%e7.20 #&#
\begin{eqnarray}
\label{contmu} \bigl\|{\mtx{X}^{(k)}}^T \mtx{V}^{(\ell)}
\bigr\|_{\ell_\infty}& =&\mathop{\max}_{i=1,\ldots,N_\ell
} \bigl\|{ \mtx{X}^{(k)}}^T
\vct{v}_i^{(\ell)}\bigr\|_{\ell_\infty}
\nonumber
\\[-8pt]
\\[-8pt]
\nonumber
& =&\mathop{
\max}_{i=1,\ldots,N_\ell} \biggl\|{\mtx{A}^{(k)}}^T{
\mtx{U}^{(k)}}^T\mtx {U}^{(\ell)}\frac{\mct{\lambda}_i^{(\ell)}}{\twonorm{\mct{\lambda
}_i^{(\ell)}}}
\biggr\|_{\ell_\infty} .
\end{eqnarray}
Now it follows from the uniform distribution of the points on each
subspace that the columns of $\mtx{A}^{(k)}$ are independently and
uniformly distributed on the unit sphere of $\R^{d_k}$. Furthermore,
the normalized dual points\footnote{Since the columns of
$\mtx{A}^{(\ell)}$ are independently and uniformly distributed on
the unit sphere of $\R^{d_\ell}$, $\mct{\lambda}_i^{(\ell)}$ in
Definition~\ref{dualpointdef} is uniquely defined with probabilty $1$.}
$\mct{\lambda}_i^{(\ell)}/\twonorm{\mct{\lambda}_i^{(\ell)}}$ are also
distributed uniformly at random on the unit sphere of
$\R^{d_\ell}$. To justify this claim, assume $\mtx{U}$ is an
orthogonal transform on $\R^{d_\ell}$ and
$\mct{\lambda}_i^{(\ell)}(\mtx{U})$ is the dual point corresponding to
$\mtx{U}\vct{a}_i$ and $\mtx{U}\mtx{A}_{-i}^{(\ell)}$. Then
%
%e7.21 #&#
\begin{equation}
\label{sim1} \mct{\lambda}_i^{(\ell)}(\mtx{U})=\mct{\lambda}
\bigl(\mtx{U}\vct{a}_i,\mtx {U}\mtx{A}_{-i}^{(\ell)}
\bigr)=\mtx{U}\mct{\lambda} \bigl(\vct{a}_i,\mtx {A}_{-i}^{(\ell)}
\bigr)=\mtx{U}\mct{\lambda}_i^{(\ell)},
\end{equation}
where we have used the fact that $\mct{\lambda}_i^{(\ell)}$ is the dual
variable in the corresponding optimization problem. On the other hand,
we know that
%
%e7.22 #&#
\begin{equation}
\label{sim2} \mct{\lambda}_i^{(\ell)}(\mtx{U})=\mct{\lambda}
\bigl(\mtx{U}\vct{a}_i,\mtx {U}\mtx{A}_{-i}^{(\ell)}
\bigr)\sim\mct{\lambda} \bigl(\vct{a}_i,\mtx{A}_{-i}^{(\ell
)}
\bigr)=\mct{\lambda}_i^{(\ell)},
\end{equation}
where $X\sim Y$ means that the random variables $X$ and $Y$ have the
same distribution. This follows from $\mtx{U}\vct{a}_i\sim\vct{a}_i$
and $\mtx{U}\mtx{A}_{-i}^{(\ell)}\sim\mtx{A}_{-i}^{(\ell)}$ since the
columns of $\mtx{A}^{(\ell)}$ are chosen uniformly at random on the
unit sphere. Combining (\ref{sim1}) and (\ref{sim2}) implies that for
any orthogonal transformation $\mtx{U}$, we have
\[
\mct{\lambda}_i^{(\ell)}\sim\mtx{U}\mct{\lambda}_i^{(\ell)},
\]
which proves the claim.

Continuing with (\ref{contmu}), since $\mct{\lambda}_i^{(\ell)}$ and
$\mtx{A}^{(k)}$ are independent, applying Lemma~\ref{infnormlog} below
with $\Delta=N_\ell L$, $N_1=N_k$, $d_1=d_k$, and $d_2=d_\ell$
gives
%
%e7.23 #&#
\[
\biggl\|{\mtx{A}^{(k)}}^T \bigl({\mtx{U}^{(k)}}^T
\mtx{U}^{(\ell)} \bigr)\frac{\mct{\lambda}_i^{(\ell)}}{\twonorm{\mct{\lambda}_i^{(\ell
)}}}\biggr\|_{\ell_\infty} \le4 \bigl(\log
\bigl[N_\ell(N_k+1) \bigr] + \log L+t \bigr)
\frac{\fronorm{{\mtx
{U}^{(k)}}^T\mtx{U}^{(\ell)}}}{\sqrt{d_k}\sqrt{d_\ell}},
\]
with probability at least $1-\frac{4}{(N_k+1){N_\ell}^2
L^2}e^{-2t}$. Finally, applying the union bound twice gives
\eqref{step2boundinco}.

%le7.5 #&#
\begin{lemma}
\label{infnormlog}
Let $\mtx{A}\in\mathbb{R}^{d_1\times N_1}$ be a matrix with columns
sampled uniformly at random from the unit sphere of
$\mathbb{R}^{d_1}$, $\mct{\lambda}\in\R^{d_2}$ be a vector sampled
uniformly at random from the unit sphere of $\R^{d_2}$ and independent
of $\mtx{A}$ and $\mct{\Sigma}\in\R^{d_1\times d_2}$ be a
deterministic matrix. For any positive constant $\Delta$, we have
\[
\bigl\|\mtx{A}^T\mct{\Sigma}\mct{\lambda}\bigr\|_{\ell_\infty} \le4 \bigl(
\log(N_1+1)+\log \Delta+t \bigr)\frac{\fronorm{\mct{\Sigma}}}{\sqrt{d_1}\sqrt {d_2}},
\]
with probability at least $1-\frac{4}{(N_1+1)\Delta^2}e^{-2t}$.
\end{lemma}
\begin{pf}
% \EJC{I would be more direct and write: $y = \langle\Sigma^T a,
% \lambda\>$. We use concentration of measure to say that $|y| \ge
% \varepsilon\|\Sigma^T a \|_2$ occurs with small probability. Then we
% bound $\|\Sigma^T a \|_2$ using Borell. }
% \MS{Given below I am not sure what you are asking me to do here.
%Should I provide a brief overview of the proof at the beginning? also
%shouldn't it be $\twonorm{\Sigma\mct{\lambda}}$}
% \EJC{I made a lot of corrections below so please check.}
The proof is standard. Without loss of generality, we assume $d_1\le
d_2$ as the other case is similar.
%N otice that since the columns of
% $\mtx{A}$ and $\mct{\lambda}$ have rotationally invariant
% distributions, we can assume that $\mct{\Sigma}$ is diagonal, with
% diagonal elements $\sigma_1\ge\sigma_2\ge\ldots\ge\sigma_{d_1}$.
To begin with, the mapping $\lambda\mapsto
\twonorm{\mct{\Sigma}\mct{\lambda}}$ is Lipschitz with constant at
most $\sigma_1$ (this is the largest singular value of
$\Sigma$). Hence, Borell's inequality gives
\[
\mathbb{P} \Bigl\{ \|\mct{\Sigma}\mct{\lambda}\|_{\ell_2}-\sqrt{
\mathbb {E} { \|\mct{\Sigma}\mct{\lambda}\|_{\ell_2}^2}}\ge
\varepsilon \Bigr\} <e^{-d_2\varepsilon^2/(2\sigma_1^2)}.
\]
Because $\mct{\lambda}$ is uniformly distributed on the unit sphere,
we have $\mathbb{E}\twonorm{\mct{\Sigma}\mct{\lambda}}^2=
\|\Sigma\|_F^2/d_2$. Plugging $\varepsilon=
(b-1)\frac{\fronorm{\mct{\Sigma}}}{\sqrt{d_2}}$ into the above
inequality, where $b=2\sqrt{\log(N_1+1)+\log
\Delta+t}$, and using $\fronorm{\mct{\Sigma}}/\sigma_1\ge1$ give
\[
\mathbb{P} \biggl( \|\mct{\Sigma}\mct{\lambda}\|_{\ell_2} >b
\frac{\fronorm{\mtx
{\Sigma}}}{\sqrt{d_2}} \biggr) \le\frac{2}{(N_1+1)^2\Delta^2}e^{-2t}.
\]

Further, letting $\vct{a}\in\R^{d_1}$ be a representative column of
$\mtx{A}$, a well-known upper bound on the area of spherical caps
gives
\[
\mathbb{P} \bigl\{\bigl|\vct{a}^T\vct{z}\bigr|>\varepsilon \|\vct{z}
\|_{\ell_2} \bigr\} \le2e^{{-d_1\varepsilon^2}/{2}}
\]
in which $\vct{z}$ is a fixed vector. We use
$\vct{z}=\mct{\Sigma}\mct{\lambda}$, and
$\varepsilon=b/\sqrt{d_1}$. Therefore, for any column $\vct{a}$ of $\mtx
{A}$ we have
\[
\mathbb{P} \biggl\{\bigl|\vct{a}^T\mct{\Sigma}\mct{\lambda}\bigr| >
\frac{b}{\sqrt{d_1}} \|\mct{\Sigma}\mct{\lambda}\|_{\ell_2} % \bigg|\twonorm{\mct{\Sigma}\mct{\lambda}}<b\frac{\fronorm{\mtx{
\biggr\}\le2e^{{-d_1\varepsilon^2}/{2}}=\frac{2}{(N_1+1)^2\Delta^2}e^{-2t}.
\]
%
% Using conditioning we get
% \begin{eqnarray}
% \mathbb{P}\bigg\{\infnorm{\vct{a}^T\mct{\Sigma}\mct{\lambda}}>\varepsilon
% &\le& \frac{2}{(N_1+1)^2\Delta^2}e^{-\frac{t}{4}}+\frac{2}{(N_1+1)^2
% &=&\frac{4}{(N_1+1)^2\Delta^2}e^{-\frac{t}{4}}.\nonumber
% \end{eqnarray}
Now applying the union bound yields
\[
\mathbb{P} \biggl( \bigl\|\mtx{A}^T\mct{\Sigma}\mct{\lambda}
\bigr\|_{\ell_\infty} > \frac
{b}{\sqrt{d_1}} \|\mct{\Sigma}\mct{\lambda}
\|_{\ell_2} \biggr)\le \frac
{2}{(N_1+1)\Delta^2}e^{-2t}.
\]
Plugging in the bound for $\twonorm{\mct{\Sigma}\mct{\lambda}}$
concludes the proof.
\end{pf}
%
%Continuing from the last few lines of the proof of Theorem~\ref{th1}
%we must prove
%For all $k=1,\ldots,\ell-1,\ell+1,\ldots,L$,\\
%For all $i=1,\ldots,N_\ell$

%we get
%(N_k+1)+\log N_\ell+ 1.5\log L\big)\frac{\fronorm{\mct{\Sigma}^{(k
%Using $\operatorname{aff}(S_k,S_\ell)=\fronorm{\mct{\Sigma}^{(k\ell)}}$ and
%applying the union bound again we get:
%k\neq\ell\bigg\}\nonumber\\
%&\ge& 1-\frac{1}{L}\sum_{k\neq\ell}^L\frac{1}{(N_k+1)N_\ell}e^{-
%s7.3 #&#
\subsection{\texorpdfstring{Proof of Theorem \protect\ref{th0}}{Proof of Theorem 1.2}}
\label{prfth0}
We prove this in two steps.
\begin{longlist}
\item[\textit{Step} 1:] We use the lower bound about the inradii used
in step 1 of the proof of Theorem~\ref{th2} with $\beta=\frac{1}{2}$, namely,
%
%e7.24 #&#
\[
\mathbb{P} \biggl\{\frac{c(\rho)}{\sqrt{2}}\sqrt{\frac{\log\rho}{d}}\le r \bigl(
\mathcal{P}^\ell_{-i} \bigr) \mbox{ for all pairs } (\ell,i)
\biggr\} \ge1-Ne^{-\sqrt{\rho}d}.
\]
\item[\textit{Step} 2:] We develop an upper bound about subspace
incoherence, namely,
%
%e7.25 #&#
\[
\mathbb{P} \biggl\{\mu(\mathcal{X}_\ell) \le\sqrt{\frac{6\log N}{n}}
\mbox{ for all } \ell \biggr\} \ge1 - \frac{2}{N}.
\]
\end{longlist}
To prove step $2$, notice that in the fully random model, the
marginal distribution of a column $\vct{x}$ is uniform on the unit
sphere. Furthermore, since the the points on each subspace are
sampled uniformly at random, the argument in the proof of Theorem
\ref{th2} asserts that the dual directions are sampled uniformly at
random on each subspace. By what we have just seen, the points
$\vct{v}_i^{(\ell)}$ are then also distributed uniformly at random on
the unit sphere (they are not independent). Last, the random
vectors $\vct{v}_i^{(\ell)}$ and $\vct{x} \in\mathcal{X} \setminus
\mathcal{X}_\ell$ are independent. The distribution of their inner
product is as if one were fixed, and applying the well-known upper
bound on the area of a spherical cap gives
\[
\mathcal{P} \biggl\{\bigl| \bigl\langle\vct{x},\vct{v}_i^{(\ell)}
\bigr\rangle\bigr|\ge\sqrt{\frac
{6\log N}{n}} \biggr\}\le\frac{2}{N^3}.
\]
Step $2$ follows by applying the union bound to at most $N^2$ such
pairs.

%s7.4 #&#
\subsection{\texorpdfstring{Proof of Theorem \protect\ref{th3}}{Proof of Theorem 2.9}}
We begin with two lemmas relating the mean and maximal value of norms
with respect to convex polytopes.
%
%le7.6 #&#
\begin{lemma}\label{product}
For a symmetric convex body in $\R^n$,
\[
\frac{M(\mathcal{K})M(\mathcal{K}^o)}{b(\mathcal{K})b(\mathcal{K}^o)} \ge\frac{1}{\sqrt{n}}.
\]
\end{lemma}
\begin{pf}
Variants of this lemma are well known in geometric functional
analysis. By definition,
\begin{eqnarray*}
\|x\|_\mathcal{K}&\le& b(\mathcal{K}) \|x\|_2,
\\
\|x\|_{\mathcal{K}^o}&\le &b \bigl(\mathcal{K}^o \bigr) \|x
\|_2,
\end{eqnarray*}
and, hence, the property of dual norms allows us to conclude that
\begin{eqnarray*}
\frac{1}{b(\mathcal{K}^o)} \|x\|_2 \le\|x\|_\mathcal{K}
&\le& b(\mathcal {K}) \|x\|_2,
\\
\frac{1}{b(\mathcal{K})} \|x\|_2 \le\|x\|_{\mathcal{K}^o} &\le& b \bigl(
\mathcal{K}^o \bigr) \|x\|_2.
\end{eqnarray*}
However, using Definition~\ref{BM}, these relationships imply that
$d(\mathcal{K},B_2^n)= b(\mathcal{K})b(\mathcal{K}^o)$.
Therefore,
\[
\frac{M(\mathcal{K})M(\mathcal{K}^o)}{b(\mathcal{K})b(\mathcal
{K}^o)}=\frac{M(\mathcal{K})M(\mathcal{K}^o)}{d(\mathcal
{K},B_2^n)}.
\]
Applying John's lemma and using Lemma~\ref{M} conclude the proof.
\end{pf}

%le7.7 #&#
\begin{lemma} \label{bound}
For a convex symmetric polytope $\mathcal{K}(\mtx{A})$, $\mtx{A}\in\R^{n\times N}$, we have
\[
n{ \biggl(\frac{M(\mathcal{K})}{b(\mathcal{K})} \biggr)}^2 \ge c \frac
{n}{\log(2N)}.
\]
\end{lemma}
\begin{pf}
By Lemma~\ref{product}, we know that
\[
\frac{M(\mathcal{K})M(\mathcal{K}^o)}{b(\mathcal{K})b(\mathcal{K}^o)} \ge\frac{1}{\sqrt{n}} \quad\Rightarrow\quad \frac{M(\mathcal{K})}{b(\mathcal{K})}\ge
\frac{1}{\sqrt{n}( {M(\mathcal{K}^o)}/{b(\mathcal{K}^o)})}.
\]
However, applying Lemma~\ref{faces} to the polytope $\mathcal{K}^o$,
which has at most $2N$ faces, gives
\[
n{ \biggl(\frac{M(\mathcal{K}^o)}{b(\mathcal{K}^o)} \biggr)}^2 \le C \log (2N) \quad\Rightarrow\quad
\frac{1}{\sqrt{n}({M(\mathcal{K}^o)}/{b(\mathcal{K}^o)})}\ge\frac
{1}{\sqrt{C\log(2N)}}.
\]
These two inequalities imply
\[
\frac{M(\mathcal{K})}{b(\mathcal{K})}\ge\frac{1}{\sqrt{C\log
(2N)}} \quad\Rightarrow \quad n{ \biggl(\frac{M(\mathcal{K})}{b(\mathcal{K})}
\biggr)}^2 \ge\frac
{1}{C}\frac{n}{\log(2N)}.
\]
\upqed\end{pf}
%
%s7.4.1 #&#
\subsubsection{\texorpdfstring{Proof of Theorem \protect\ref{th3} [part \textup{(a)}]}{Proof of Theorem 2.9 [part (a)]}}
The proof is in two steps: %First, we will show that for an inlier (a
%point on one of the subspaces) the optimal value of the optimization
%problem (\ref{eqel2}) is upper bounded by $\underset{\ell,i}{\max}
%outlier point the optimal value of the optimization problem (
%proves theorem~\ref{th3} part (a).

%
\begin{longlist}[(1)]
\item[(1)] For every inlier point $\vct{x}_i^{(\ell)}$,
%
%e7.26 #&#
\begin{equation}
\label{step1inlier} \operatorname{optval} \bigl(\vct{x}_i^{(\ell)},
\mtx{X}_{-i} \bigr)\le\frac
{1}{r(\mathcal{P}_{-i}^{\ell})}.
\end{equation}
\item[(2)] For every outlier point $\vct{x}_i^{(0)}$, with probability
at least $1-e^{-c{nt^2}/{\log N}}$, we have
\[
(1-t)\frac{\lambda(\gamma)}{\sqrt{e}}\sqrt{n}\le\operatorname{optval} \bigl(\vct
{x}_i^{(0)},\mtx{X}_{-i} \bigr).
\]
\end{longlist}
%
%pa7.4.1.1 #&#
\paragraph*{Proof of step 1}
%
%le7.8 #&#
\begin{lemma}
\label{step1}
Suppose $\vct{y}\in\operatorname{Range}(\mtx{A})$, then
\[
\operatorname{{optval}} (\vct{y},\mtx{A} )\le\frac{\twonorm{\vct
{y}}}{r(\mathcal{K}(\mtx{A}))}.
\]
\end{lemma}
\begin{pf}
As stated before,
%
%e7.27 #&#
\[
\label{dualform} \operatorname{optval} (\vct{y},\mtx{A} )={\|\vct{y}
\|}_{\mathcal{K}(\mtx
{A})}.
\]
Put $\mathcal{K}(\mtx{A}) = \mathcal{K}$ for short. Using the
definition of the max norm and circumradius,
%
%e7.28 #&#
\begin{equation}
\|\vct{y}\|_{\mathcal{K}} = \|\vct{y}\|_{\ell_2} \biggl\Vert\frac{\vct
{y}}{\twonorm{\vct{y}}}
\biggr\Vert_{\mathcal{K}} \le \|\vct {y}\|_{\ell_2} b(\mathcal{K})= \|\vct{y}
\|_{\ell_2} R \bigl( \mathcal{K}^o \bigr)= \frac{\twonorm
{\vct{y}}}{r(\mathcal{K})}.
\end{equation}
The last equality follows from the fact that maximal norm on the unit
sphere and the inradius are the inverse of one another (Lemma
\ref{invnormlemma}).\vadjust{\goodbreak}
\end{pf}
Notice that
\[
\operatorname{optval} \bigl(\vct{x}_i^{(\ell)},
\mtx{X}_{-i} \bigr)\le\operatorname {optval} \bigl(\vct{x}_i^{(\ell)},
\mtx{X}_{-i}^{(\ell)} \bigr),
\]
and since $\twonorm{\vct{x}_i^{(\ell)}}=1$, applying the above lemma
with $\vct{y}=\vct{x}_i^{(\ell)}$ and $\mtx{A}=\mtx{X}_{-i}^{(\ell)}$ gives
%
%e7.29 #&#
\[
\operatorname{optval} \bigl(\vct{x}_i^{(\ell)},
\mtx{X}_{-i}^{(\ell)} \bigr)\le\frac
{1}{r(\mathcal{P}_{-i}^{\ell})}.
\]
Combining these two identities establishes \eqref{step1inlier}.

%pa7.4.1.2 #&#
\paragraph*{Proof of step 2}
We are interested in lower bounding $\operatorname{optval} (\vct{y},\mtx
{A} )$ in which $\mtx{A}$ is a fixed matrix and $\vct{y}\in\R^n$ is
chosen uniformly at random on the unit sphere. Our strategy consists in
finding a lower bound in expectation, and then using a concentration
argument to derive a bound that holds with high probability.

%le7.9 #&#
\begin{lemma}[(Lower bound in expectation)]\label{lowerbound}Suppose $\vct
{y}\in\mathbb{R}^n$ is a point chosen uniformly at random on the unit
sphere and $\mtx{A}\in\mathbb{R}^{n\times N}$ is a matrix with
unit-norm columns.
Then
\[
\mathbb{E} \bigl\{\operatorname{{optval}} (\vct{y},\mtx{A} ) \bigr\}>
\cases{ \displaystyle\frac{1}{\sqrt{e}}\sqrt{\frac{2}{\pi}}
\frac{n}{\sqrt{N}}, & \quad $\mbox{if } 1\le\displaystyle\frac{N}{n}\le e,$ \vspace*{2pt}
\cr
\displaystyle\frac{1}{\sqrt{e}}\sqrt{\frac
{2}{\pi e}}\sqrt{\frac{n}{\log\frac{N}{n} }}, &\quad  $\mbox{if
}  \displaystyle\frac{N}{n}\ge e.$ } %
\]
\end{lemma}
\begin{pf}
Since $\mbox{optval} (\vct{y},\mtx{A} )={\|\vct{y}\|}_{\mathcal
{K}(\mtx{A})}$, the expected value is equal to $M^*(\mathcal
{K}^o)=M(\mathcal{K})$. Applying Urysohn's theorem (Theorem \ref
{Urysohn}) gives
\[
\label{meanwidthUrysohn} M^* \bigl(\mathcal{K}^o \bigr) \ge{ \biggl(
\frac{\operatorname{vol}(\mathcal{K}^o)}{\operatorname
{vol}(B_2^n)} \biggr)}^{{1}/{n}}.
\]
It is well known that the volume of the $n$-dimensional sphere with
radius one is given by
\[
\operatorname{vol} \bigl(B_2^n \bigr)=
\frac{\pi^{n/2}}{\Gamma(({n}/{2})+1)}.
\]
The well-known Stirling approximation gives
\[
\Gamma \biggl(\frac{n}{2}+1 \biggr)\ge\sqrt{2\pi}e^{-n/2}{
\biggl(\frac{n}{2} \biggr)}^{(n+1)/2},
\]
and, therefore, the volume obeys
\[
\label{volsphere} \operatorname{vol} \bigl(B_2^n \bigr)
\le{ \biggl(\sqrt{\frac{2\pi e}{n}} \biggr)}^n.
\]
Note that if $\{\vct{a}_i\}_{i=1}^N$ is a family of $n$-dimensional
unit-norm vectors, then for $p\ge1$,
\[
{ \Biggl(\frac{1}{n}\sum_{i=1}^{N}
\abs{\vct{a}_i}^p \Biggr)}^{{1}/{p}}\le { \biggl(
\frac{N}{n} \biggr)}^{{1}/{p}}.
\]
Applying Lemma~\ref{Ball} for $p\ge2$ gives
\[
{\operatorname{vol} \bigl(\mathcal{K}^o \bigr)}^{{1}/{n}}\ge
\frac{2\sqrt{2}}{\sqrt {p}{ ({N}/{n} )}^{1}/{p}} .
\]
The right-hand side is maximum when $p=2\log\frac{N}{n}$, which is
larger than $2$ as long as $\frac{N}{n}\ge e$. When $\frac{N}{n}< e$,
we shall use $p=2$. Plugging in this value of $p$ in the bound of Lemma
\ref{Ball}, we conclude that
\[
{\operatorname{vol} \bigl(\mathcal{K}^o \bigr)}^{{1}/{n}}\ge
\cases{\displaystyle \frac{2}{\sqrt{\frac{N}{n}}}, &\quad $ \mbox{if } 1\le\displaystyle\frac
{N}{n}\le
e,$ \vspace*{2pt}
\cr
\displaystyle\frac{2}{\sqrt{e}} \frac{1}{\sqrt{\log\frac{N}{n}} }, & \quad$\mbox{if }     \displaystyle\frac{N}{n}\ge e.$} %
\]
Finally, this idenitity together with the approximation of the volume
of the sphere conclude the proof.\vspace*{-2pt}
\end{pf}
%
%le7.10 #&#
\begin{lemma}[(Concentration around mean)]In the setup of Lemma~\ref{lowerbound},
%
%e7.30 #&#
\[
\operatorname{{optval}}(\vct{y},\mtx{A})\ge(1-t)\mathbb{E} \bigl\{
\operatorname{optval}(\vct{y},\mtx{A}) \bigr\},
\]
with probability at least $1-e^{-c{nt^2}/{\log(2N)}}$.\vspace*{-2pt}
\end{lemma}
\begin{pf}
The proof follows from Theorem~\ref{concentration} and applying Lemma
\ref{bound}.\vspace*{-2pt}~%
\end{pf}
These two lemmas (Lower bound in expected value and Concentration
around mean) combined with the union bound give the first part of
Theorem~\ref{th3}.\vspace*{-2pt}
%s7.4.2 #&#
\subsubsection{\texorpdfstring{Proof of Theorem \protect\ref{th3} part (b)}{Proof of Theorem 2.9 part (b)}}
This part follows from the combination of the proof of Theorem
\ref{th3} part (a) with the bound given for the inradius presented in
the proof of Theorem~\ref{th2}.\vspace*{-2pt}

\subsection{\texorpdfstring{Proof of Theorem \protect\ref{th0outlier}}{Proof of Theorem 1.3}}
The proof follows Theorem~\ref{th3} with $t$ a small number. Here we
use $t=1-\frac{1}{\sqrt{2}}$.

\section*{Acknowledgments}
E. J. Cand\'es would like to thank Trevor Hastie for discussions related to
this paper. M. Soltanolkotabi acknowledges fruitful conversations with Yaniv Plan,
and thanks Gilad Lerman for clarifying some of his results and Ehsan
Elhamifar for comments on a previous draft. We are grateful to the
reviewers for suggesting new experiments and helpful comments.\vadjust{\goodbreak}

% imsref loaded by akundreckaite, 2012-10-12 10:51:50
% imsref loaded by akundreckaite, 2012-10-12 13:02:45
% imsref loaded by akundreckaite, 2012-11-19 10:16:54
% imsref loaded by akundreckaite, 2012-11-19 10:24:38
% imsref loaded by akundreckaite, 2012-11-19 10:26:15

%

%suskaldyti doi

\printaddresses

\end{document}